\documentclass[%
 reprint,
superscriptaddress,
 amsmath,amssymb,
 aps,
]{revtex4-1}

\usepackage{graphicx}
\usepackage[caption=false]{subfig}
\usepackage{soul}
\usepackage{xcolor}
\usepackage{dcolumn}
\usepackage{bm}
\usepackage[colorlinks=true, pdfstartview=FitV, linkcolor=red,
citecolor=blue, urlcolor=blue]{hyperref}

\def\pmin{$p_{\rm min}^{\rm J}$}

\begin{document}

\preprint{APS/123-QED}

\title{Mini-jet quenching in a concurrent jet+hydro evolution and the non-equilibrium quark-gluon plasma}

\author{Daniel Pablos}
 \email{daniel.pablos.alfonso@to.infn.it}
\affiliation{%
 INFN, Sezione di Torino, via Pietro Giuria 1, I-10125 Torino, Italy
}%

\author{Mayank Singh}
\affiliation{School of Physics \& Astronomy, University of Minnesota, Minneapolis, MN 55455, USA}
\affiliation{
 Department of Physics, McGill University, 3600 University Street, Montr\'eal, QC, H3A 2T8, Canada\\
}%

\author{Sangyong Jeon}
\affiliation{
 Department of Physics, McGill University, 3600 University Street, Montr\'eal, QC, H3A 2T8, Canada\\
}%

\author{Charles Gale}
\affiliation{
 Department of Physics, McGill University, 3600 University Street, Montr\'eal, QC, H3A 2T8, Canada\\
}%

\newcommand{\rmd}{{\rm d}}
\newcommand{\rme}{{\rm e}}

\renewcommand{\arraystretch}{1.4}

\date{\today}

\begin{abstract}
Mini-jets, created by perturbative hard QCD collisions at moderate energies, can represent a significant portion of the total multiplicity of a heavy-ion collision event. Since their transverse momenta are initially larger than the typical saturation scale describing the bulk of the equilibrating QGP, they ought to be described through the physics of parton energy loss. Indeed, their typical stopping distances are larger than the usual hydrodynamization time, so they do not in general hydrodynamize at the same pace than the bulk of the collision. Therefore, in general mini-jets cannot be described solely by a unique pre-equilibrium stage that bridges the initial, over-occupied glasma state, with the hydrodynamical evolution. In this work we make use of a new concurrent mini-jet+hydrodynamic framework in which the properties of the hydrodynamically evolving QGP are modified due to the injection of energy and momentum from the mini-jets. We study the system for different choices of the minimum transverse momentum associated to mini-jet production. In order to achieve a realistic description of charged particle multiplicity, the amount of entropy associated to the low-$x$ initial state needs to be reduced. Moreover, the fact that the injected momentum from the randomly oriented mini-jets is not correlated with the spatial gradients of the system reduces overall flow, and the value of the QGP transport coefficients needs to be reduced accordingly in order to describe the measured flow coefficients in experiments. They are, in effect, an important new source of fluctuations, resulting in a spikier, notably modified hydrodynamical evolution when compared to the scenario in which the presence of mini-jets is ignored. We avow that their abundance makes it necessary to include their physics in holistic descriptions of heavy-ion collisions. We discuss the impact of the mini-jets on a number of observables, such as $p_T$ spectra and $p_T$-differential flow $v_n$ for a wide range of centrality classes. In contrast to elliptic, triangular or quadrangular flow, we find that directed flow, $v_1$, has the strongest potential to discriminate between different mini-jet production rates.
\end{abstract}

\maketitle


\section{Introduction}

Heavy-ion collisions in modern accelerators have succeeded in reproducing the extreme conditions that existed immediately after the Big Bang. In such an environment, ordinary matter becomes the most perfect fluid ever measured in Nature: the quark-gluon plasma (QGP) \cite{Baym:2016wox,Busza:2018rrf}. Relativistic hydrodynamic simulations have been successfully applied to describe the strong correlations among the tens of thousands of particles that fly to the detectors. For a few fm/c, the system is well described by the hydrodynamic explosion of a liquid droplet of deconfined QCD matter \cite{Heinz:2013th,Gale:2013da}.

Why exactly hydrodynamics works this well in describing a system which spends a large part of its evolution far from local equilibrium, and where gradients are not small \cite{Niemi:2014wta}, is currently under very active investigation \cite{Schlichting:2019abc,Soloviev:2021lhs}. Recent developments based on the bottom-up thermalization scenario \cite{Baier:2000sb,Kurkela:2014tea} show that QCD effective kinetic theory (EKT) manifests hydrodynamic behaviour after a characteristic time $\tau_R$, provided that it is smaller than the system size $R$~\cite{Kurkela:2018vqr,Kurkela:2018wud}. This behaviour is mostly driven by the radiative break-up of the large number of gluons that make up the initially over-occupied system called the Color Glass Condensate (CGC) \cite{Iancu:2003xm}. 

The collision terms that enter the EKT equations have been derived for a weak-coupling QCD system based on a quasi-particle picture \cite{Arnold:2002zm,Arnold:2002ja}. Such elastic and inelastic processes have also been used in the description of the energy loss  that high-energy jets suffer while traversing the QGP \cite{Schenke:2009gb,Du:2020dvp}. The set of modifications experienced by energetic jets due to their passage through the medium are commonly known as jet quenching \cite{dEnterria:2009xfs,Majumder:2010qh,Mehtar-Tani:2013pia}. In turn, the passage of the energetic jets through the medium has been found to modify the QGP background as well~\cite{Tachibana:2019hrn,Cao:2020wlm,Luo:2021iay}. A large number of studies have shown that many of the jet quenching observables measured in experiments are best understood with a proper treatment of energy-momentum conservation through the consideration of the fate of the lost energy and medium back-reaction \cite{He:2015pra,Casalderrey-Solana:2016jvj,Tachibana:2017syd,Milhano:2017nzm,KunnawalkamElayavalli:2017hxo,Chen:2017zte,He:2018xjv,Park:2018acg,Casalderrey-Solana:2019ubu,Chang:2019sae,Pablos:2019ngg,Chen:2020tbl,Yang:2021qtl,He:2022evt}. The radiative break-up and further re-scattering of such hard probes eventually leads to the hydrodynamization of part of the jet energy, which shows up in observables as an excess of soft particles at large angles with respect to the jet axis. 

High energy QCD processes are rare, most of them producing at most a single energetic dijet pair per central heavy-ion collision (e.g. for jets with $p_T\geq 20$ GeV at $\sqrt{s}=2.76$ ATeV). This is not the case for the production rate of lower $p_T$ jets, the so-called mini-jets, which are produced abundantly across all collision centralities. They are mini parton showers that experience the same kind of processes embedded in the EKT approach used to describe hydrodynamization of energy both at low and high $p_T$. Their energy range bridges the gap between the physics of the bottom-up scenario and that of jet quenching: hard enough not to become part of the bulk of the system at the same pace as the lowest $p_T$ quanta, and soft enough to have to consider the simultaneous propagation of a large number of such mini-jet pairs instead of just a single pair. Therefore, these lower $p_T$ jets can no longer be considered as mere probes of the system, but rather as a sizeable part of it. 

Previous work addressing the impact of the presence of a non-equilibrated sector along with an equilibrated one have focused on transverse momentum and flow correlations~\cite{Pang:2009zm,Werner:2013tya,Andrade:2014swa,Schulc:2014jma,Crkovska:2016flo,Okai:2017ofp,Bravina:2020sbz,Zhao:2021vmu,Ryu:2021qwq} as well as strangeness production~\cite{Becattini:2008ya,Kanakubo:2018vkl}. The present work follows up this idea with the establishment of a novel framework which utilizes validated physical models that simulate the evolution of the different stages of a heavy-ion collision, for both the soft and hard sectors. By connecting hydrodynamics and jet quenching, we present the phenomenological aspects that emphasize the importance of working within a holistic description of heavy-ion collisions.

One of the key aspects of the influence of the presence of mini-jets on the bulk of the system is their random angular orientation in the transverse plane at production time $\tau=0$. Given that in the bulk of the system flow develops along the pressure gradients that originate due to the spatial anisotropies of the energy density profile, as dictated by hydrodynamics, the fact that a sizeable part of the system energy is initially uncorrelated to such spatial deformations will in general tend to dilute collective flow \cite{Schulc:2014jma,Okai:2017ofp}. Considering the strong effect that transport coefficients, specially shear viscosity, have in the modulation of collective flow, one expects that the presence of mini-jets will alter the phenomenological extraction of such transport coefficients through comparison with data.

The fact that the mini-jets have on average a larger energy than the rest of the quanta of the bulk has two important consequences. On one hand, it means that it does not suffice to consider a single hydrodynamization time for the whole system, typically around $\tau_0 \approx 1$ fm/c, as some of the mini-jets might retain a sizeable part of their energy well above 3 fm/c. On the other hand, having sectors of the system that hydrodynamize at different paces implies that the local properties of the medium perceived by a given mini-jet, or by a given hydrodynamic cell, depend on the amount of energy and momentum deposited in the causal past by other mini-jets. 
These aspects call for the development of a fully concurrent jet-hydrodynamical evolution, in which the hydrodynamical profile is updated at each time step through the source terms injected by the mini-jets.

The rest of this paper is organized as follows: in Section~\ref{sec:quenching} we perform some estimates on the stopping distances of mid-$p_T$ partons, while Section~\ref{sec:model} is devoted to the description of the model used in this work. Results on the modified hydrodynamical evolution and some observables are presented in Section~\ref{sec:results}. Finally, we summarize our findings and look ahead in Section~\ref{sec:summary}.

\section{(Mini-)Jet quenching}
\label{sec:quenching}

\begin{figure}[t]
    \centering
    \includegraphics[width=0.45\textwidth]{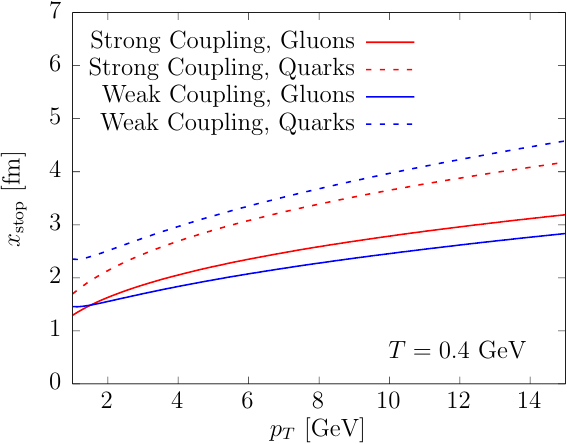}
    \caption{Stopping distance $x_{\rm stop}$ as a function of gluon $p_T$, with $T=0.4$ GeV.}
    \label{fig:timestop}
\end{figure}

Partons with energy $E \gg T$ take a finite time to reach the thermal scale, when $E \approx T$, after traversing the so called stopping distance $x_{\rm stop}$. In a weakly-coupled QGP, it has been shown that to leading logarithmic approximation (LL), when $\textrm{ln}(E/T)$ is large, the stopping distance goes like \cite{Arnold:2009ik}
\begin{equation}
\label{eq:weakstop}
    x_{\rm stop}^{\rm {pQCD}}=\frac{1}{a_i \alpha_s^2 T}\sqrt{\frac{E/T}{\textrm{ln}(E/T)}}\, , 
\end{equation}
where $\alpha_s$ is the (fixed) strong coupling constant, $T$ is the QGP temperature and $a_i$ is a species dependent parameter. The characteristic energy scaling of $E^{1/2}$ is due to the Landau-Pomeranchuk-Migdal (LPM) effect, in which successive collisions with the medium during the formation time of the induced emission lead to destructive interferences, as properly taken into account in the BDMPS-Z~\cite{Baier:1996kr,Baier:1996sk,Zakharov:1996fv,Zakharov:1997uu} and AMY~\cite{Arnold:2002zm,Arnold:2002ja} energy loss rates, from which Eq.\eqref{eq:weakstop} can be derived. 

The stopping distance can also be defined at strong coupling, in the non-perturbative regime, by using the holographic duals of certain supersymmetric theories \cite{Chesler:2008uy}. It has been shown that the maximum stopping distance of a light quark in a strongly-coupled plasma scales as \cite{Chesler:2008uy,Gubser:2008as,Hatta:2008tx}
\begin{equation}
\label{eq:strongstop}
    x_{\rm stop}^{\rm {AdS/CFT}}=\frac{1}{\kappa_i T} \left(\frac{E}{T}\right)^{1/3} \, ,
\end{equation}
where $\kappa_i$ is a species dependent parameter. Discussing in terms of stopping distances is convenient in this case, since comparing bremmstrahlung rates, for instance, would not be a well posed question given the absence of the notion of individual, perturbative quanta at strong coupling.

We show in Fig.~\ref{fig:timestop} a comparison of the stopping distances between weak coupling and strong coupling scenarios, for quarks and gluons, for educated choices of the parameters. For the weak coupling result of Eq.~\eqref{eq:weakstop}, we take the values of $a_i$ computed in \cite{Arnold:2009ik} at LL, namely $a_q \simeq 3.9$ and $a_g \simeq 6.3$ (to the level of precision needed in this discussion and the chosen energy range, it is not relevant whether we use the low $E$ or high $E$ results). The value of the strong coupling constant is fixed to $\alpha_s=0.3$, as it is customary in pQCD jet quenching phenomenology \cite{Schenke:2009gb}. Regarding the AdS/CFT result of Eq.~\eqref{eq:strongstop}, the stopping distance of a gluon is reduced by a factor $(C_A/C_F)^{1/3}$ compared to that of a quark \cite{Gubser:2008as}, which leads to $\kappa_g=(C_A/C_F)^{1/3} \, \kappa_q$, where $\kappa_q=0.4$, as extracted from phenomenological studies of jet quenching at strong coupling \cite{Casalderrey-Solana:2018wrw}. We assume the average initial temperature of the QGP at LHC to be $T=0.4$ GeV. The important message from Fig.~\ref{fig:timestop} is that for reasonable values of the parameters, even fairly low energy partons take a sizeable amount of time to thermalize, considerably more than the usual hydro starting time of $\tau_0\approx 0.4$ fm/c. Moreover, even the pre-equilibrium stage matching time with hydrodynamics of around $\tau_{\rm pre}\approx 1$ fm/c would not be enough to account for the dynamics of the mid-energy gluons with $p_T \geq 5$ GeV. 

\section{A concurrent mini-jet+hydrodynamics evolution}
\label{sec:model}

\begin{figure}[t!]
    \centering
    \includegraphics[width=0.45\textwidth]{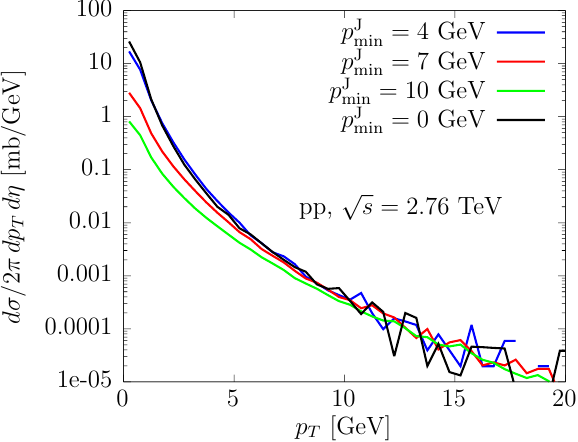}
    \caption{Charged hadron production in proton-proton collisions at $\sqrt{s}=2.76$ TeV, within $|\eta|<2$, for different values of \pmin$>$ 0 with hard QCD processes only, and also including diffractive and non-diffractive processes in the minimum bias setup symbolically represented by \pmin= 0.}
    \label{fig:crossptmin}
\end{figure}

\begin{figure}[t!]
    \centering
    \includegraphics[width=0.49\textwidth]{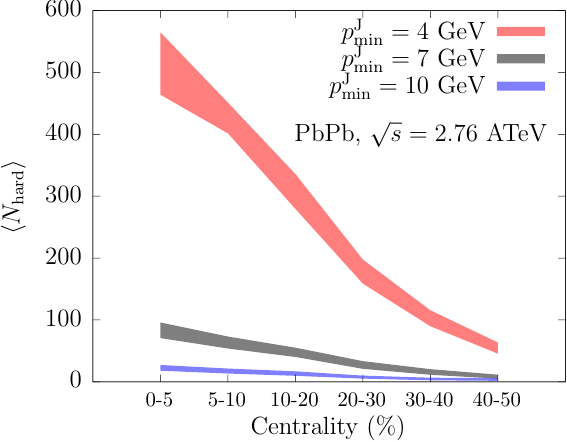}
    \caption{Average number of hard collisions as a function of centrality for different choices of \pmin.}
    \label{fig:nhard}
\end{figure}

In this work we assume that the initial state of heavy-ion collisions at low $x$ is effectively described by the CGC effective theory, governed by the scale $Q_s$. Harder QCD processes, such as mini-jet production, are assumed to decouple from the physics of the condensate inasmuch as the minimum $p_T$ of the initial back-to-back parton pair, \pmin, is chosen to be greater than $Q_s$. 
It is currently not known how to model the evolution of the lowest $x$ components of the system energy together with the harder modes corresponding to the mini-jets within a single approach. 
Other related works have adopted different sets of prescriptions to gauge \pmin, such as those based on NLO mini-jet production computations \cite{Paatelainen:2013eea,Niemi:2015qia}, survival probabilities within a hot medium \cite{Werner:2013tya,Okai:2017ofp}, or directly fixed parameter values \cite{Lokhtin:2008xi,Zhao:2021vmu}.
For the purpose of this work, \pmin is regarded as a free parameter that we vary in order to study the impact of the presence of the mini-jets as a function of their abundance, increasing (decreasing) with decreasing (increasing) \pmin.

In nucleon-nucleon collisions, particle production at high $p_T$ is dominated by hard QCD processes, while at lower $p_T$ diffractive and non-diffractive processes play a more dominant role. Our current focus are the mini-jets produced by the hard QCD processes only. By choosing a given \pmin, we will in general get an incomplete description of the spectra below such value, i.e. when $p_T<$ \pmin. That is fine if the system has other mechanisms of producing particles at these low $p_T$, as it is the case in heavy-ion collisions. However, in the limit in which the nuclear overlap, or system size is greatly reduced, one should recover the nucleon-nucleon scenario. From this point of view, it would be desirable to be able to describe particle production as inclusively as possible, which motivates the adoption of a relatively low \pmin. The comparison of charged hadron production within PYTHIA for different choices of \pmin is performed in Fig.~\ref{fig:crossptmin}. As anticipated, spectra coincide only at high enough $p_T$, and the higher \pmin yields the lower production at low $p_T$. The choice of \pmin= 0 actually corresponds to the minimum bias setup in PYTHIA (it does not simply refer to the diverging limit $p_T \rightarrow 0$ of pQCD hard scattering cross section in PYTHIA)  where soft QCD diffractive and non-diffractive processes need to be included. While the choice of \pmin= 0 would certainly lead to double counting when simultaneously including the CGC phase, the values \pmin= 4, 7, 10 GeV, which will be used throughout this work, are hard QCD processes which are assumed to be decoupled from the rest of the system at collision time.  Note that it is likely that the actual value of \pmin will be centrality/multiplicity-dependent \cite{Zhao:2021vmu}. We do not explore this added level of complication in the current study.

We describe the over-occupied system at low $x$ using the IP-Glasma model. In this model, each approaching nucleus is represented by a strong gluon color field (small $x$ partons) generated by color charges on one of the light-cone axes (large $x$ partons).
The initial stage of the collision is then represented by the interaction of the two strong gluon fields inside the forward light-cone \cite{Schenke:2012wb,McDonald:2016vlt}.
We introduce the parameter $s_{\rm factor}$ which multiplies the energy density at the moment of matching with hydro; its value is adjusted to reproduce the final charged particle multiplicity. 

Within the heavy-ion environment, we produce mini-jets using PYTHIA8 \cite{Sjostrand:2006za,Sjostrand:2007gs}, setting a minimum scale for production of $\widehat{p_T}^{\rm min}=$ \pmin and using nuclear parton distribution functions (nPDFs) as parametrized by EPS09 \cite{Eskola:2009uj} at leading order (LO). Only hard QCD events are considered. At the location of each of the binary collisions 
calculated within
the IP-Glasma model, we accept the creation of a given mini-jet pair with probability $\sigma_i/\sigma_{NN}$, where $\sigma_i$ is the cross section for the candidate hard QCD process and $\sigma_{NN}$ is the total nucleon-nucleon cross section. In this work we present results for $\sqrt{s}=2.76$ ATeV, for which we take $\sigma_{NN}=64$ mb \cite{Loizides:2016djv}.

Because of the steeply falling jet spectrum, the number of minijets produced wildly varies as a function of \pmin, as can be seen in Fig.~\ref{fig:nhard}. We show the average number of hard QCD collisions with momentum transfers $p_T>$ \pmin as a function of centrality, for different choices of \pmin. Due to the power-law behaviour of the jet spectrum, differences among \pmin choices are quite large. Having \pmin= 4 GeV, a value which satisfies \pmin$> Q_s$,  leads to a strikingly high number of minijets produced in central collisions. In this model, the evolution with centrality is only due to the fewer number of binary collisions, $N_{\rm coll}$, when one moves towards more peripheral events.

The space-time structure and energy loss dynamics of mini-jets are in this work those used in the hybrid strong/weak coupling model \cite{Casalderrey-Solana:2014bpa,Casalderrey-Solana:2015vaa}, in which parton splittings are treated perturbatively using DGLAP evolution and the interaction with the QGP is described at strong coupling. Each parton propagates through the medium for its formation time, estimated as $\tau_f=2E/Q^2$, where $E$ is the parton energy and $Q$ its virtuality. The rate at which energy and momentum are being transferred from the hard mini-jet modes into the long wavelength hydrodynamic modes has been computed using holography for $\mathcal{N}=4$ SYM at large coupling with large $N_c$, and reads~\cite{Chesler:2014jva,Chesler:2015nqz}
\begin{equation}
\label{eq:elossrate}
   \left. \frac{\rmd E}{\rmd x}\right|_{\rm strongly~coupled}= - \frac{4}{\pi} E_{\rm in} \frac{x^2}{x_{\rm stop}^2} \frac{1}{\sqrt{x_{\rm stop}^2-x^2}} \quad ,
\end{equation}
where $E_{\rm in}$ is the initial energy of the parton and $x_{\rm stop}$ has already been defined in Eq.~\eqref{eq:strongstop}. Note that, in contrast to what it is customary in weakly coupled pictures of the jet/medium interaction, here we do not need to define an energy scale below which one considers a parton to be hydrodynamized, or thermal -- Eq.~\eqref{eq:elossrate} is precisely the amount of hydrodynamized jet energy per unit length. After each time step, the energy and momentum lost by the jet are injected in the hydrodynamical system via a source term, which we define below. For simplicity, we ignore energy loss between the initial time $\tau=0$ and the hydrodynamical starting time chosen in this work, $\tau_0=0.4$ fm/c. This results into a slight overestimate of the time it takes for a parton to get stopped in the plasma \footnote{We have checked the impact of this choice by re-running some of our results assuming that the temperature profile up to $\tau_0$ is exactly the one at $\tau=0$, and quenching the mini-jets accordingly during such time (the accumulated energy and momentum lost from the jets during this time is injected all at once at $\tau_0$). All the observables we have checked are largely insensitive to either this or our default setup.}. This can be improved in the future, for instance including recent results on jet energy loss in the pre-equilibrium Glasma phase \cite{Ipp:2020nfu,Ipp:2020mjc,Carrington:2021dvw,Carrington:2022bnv}. We also turn off energy loss for those partons sitting on fluid cells with a temperature below the pseudo-critical temperature $T_c$ (although further interactions will be allowed in the hadron resonance gas phase).

The hydrodynamical evolution, after being matched to the Glasma phase at time $\tau_0$, is described using MUSIC~\cite{Schenke:2010nt,Schenke:2010rr,Paquet:2015lta}, which solves the relativistic hydrodynamic equations of motion including both shear~\cite{Schenke:2010rr} and bulk~\cite{Ryu:2015vwa} viscous corrections. The shear viscosity over entropy density parameter, $\eta/s$, is taken to be a constant that we will adjust for different values of \pmin. A more general parametrization that accounts for its temperature dependence~\cite{Niemi:2011ix} will be studied in future work. Regarding the bulk viscosity over entropy density parameter, $\zeta/s$, we follow the parametrization of its temperature dependence presented in Ref.~\cite{Denicol:2009am}. In this work we do not attempt at varying $\zeta/s$, leaving it for future multi-parameter fit studies. We employ the equation of state computed by the HotQCD Collaboration ~\cite{HotQCD:2014kol}. Energy and momentum injected by the mini-jets enters the hydrodynamic equations of motion via
\begin{equation}
    \partial_{\mu}T^{\mu \nu}_{\rm hydro} = J^{\nu},
\end{equation}
where the source term $J^{\nu}$ is parametrized as a Gaussian with a width $\sigma_x$ in both the $x$ and $y$ transverse directions and a width $\sigma_{\eta}$ in pseudo-rapidity,
\begin{equation}
\label{eq:source}
J^{\nu}=\sum_i \frac{\Delta P_i^{\nu} }{\Delta \tau (2 \pi)^{3/2} \sigma_x^2 \sigma_{\eta}\tau} e^{-\frac{\Delta x_i^2 +\Delta y_i^2}{2 \sigma_x^2} } e^{-\frac{\Delta \eta_i^2}{2 \sigma_{\eta}^2} } \, ,
\end{equation}
where $\Delta \tau$ is the evolution time step.
In the last expression, the sum $i$ runs over all four-momentum depositions $\Delta P^{\nu}$ at time $\tau$ which occurred at a distance $\Delta x$, $\Delta y$ and $\Delta \eta$ from the local fluid cell where the source term is evaluated. $\Delta P^{\nu}$ is determined for each propagating parton according to the hydrodynamization rate Eq.~\eqref{eq:elossrate}, where it is assumed that momentum deposition happens along the parton orientation. In the present work we use $\sigma_x=0.4/\sqrt{2}$ fm and $\sigma_{\eta}=0.4/\sqrt{2}$, and choose to cut away the contributions coming from the tails of the Gaussian that are beyond $5 \sigma$ in any direction. The study of the dependence of our results on the parametrization of the source term, as well as the introduction of a causal source term based on the relativistic causal diffusion equation~\cite{Aziz:2004qu,JETSCAPE:2020uew,Tachibana:2020mtb} will be done in future work.

Fluid cells are converted into particles through the usual Cooper-Frye procedure~\cite{Cooper:1974mv}, where the multiplicity of each hadron species is sampled for each fluid element belonging to an isothermal freeze-out hypersurface at $T_{\rm switch}=T_c=145$ MeV. The used viscous corrections to the phase space distribution, following~\cite{Schenke:2020mbo}, are from the 14-moment approach for the shear $\delta f$ and from the Chapman-Enskog form for the bulk $\delta f$~\cite{Ryu:2015vwa,Schenke:2020mbo}.

In order to hadronize the surviving mini-jet partons, those that did not completely hydrodynamize, we use the Lund string model \cite{Andersson:1983ia} present in PYTHIA8. This model requires to have sets of colour neutral partonic systems. We achieve these via two different mechanisms, depending on whether a jet parton has crossed the freezeout hypersurface: local thermal color neutralization (LTCN) and corona color neutralisation (CCN). The combination of these two methods of color neutralization ensures that partons that have not been quenched hadronize independently from the medium, via CCN, preserving to a large extent the vacuum-like features an unquenched jet should have~\cite{Kumar:2019bvr}. On the other hand, partons whose color has been randomized due to medium interactions will have a modified colour flow and consequently are likely to hadronize together with the thermal degrees of freedom within the QGP, via LTCN.

We discuss LTCN first. We record the space-time position of the last time that a surviving jet parton crosses the freezeout hypersurface, $x_{\rm \tiny FO}^{\mu}$. Crossing the freezeout hypersurface is defined as being at a fluid cell with $T>T_c$ at a given time step and at a fluid cell with $T<T_c$ at the next time step. A parton that from time $\tau_0$ is never found in a fluid cell with $T>T_c$ is hadronized via CCN. If $x_{\rm \tiny FO}^{\mu}$ exists, we look for a nearby fluid cell, where ``nearby'' means that it is within 2 times the simulation step size of the 4 space-time variables, i.e. $2 \delta x$, $2 \delta y$, $2 \delta \eta$ and $2 \delta \tau$, from $x_{\rm \tiny FO}^{\mu}$. 
We compute the particle spectrum associated to the first nearby cell found, differential in $p_T$, $\phi$ and $\eta$, for quarks with mass $m=0.33$ GeV. In order to achieve a color neutral object, we use this spectrum to sample the kinematics of one thermal quark if the jet parton is a quark, or two thermal quarks if the jet parton is a gluon. While we sample $u$ or $d$ quarks with equal probability (ignoring $s$ quarks in the present work), the particle or anti-particle nature of the thermal quarks is adjusted to ensure colour neutrality. 
The physics of recombination, partially captured by our LTCN implementation of hadronization, has been shown to be important to describe particle ratio and spectra in the intermediate $p_T$ region~\cite{Zhao:2020wcd,Tjemsland:2022dzz,Han:2016uhh,Zhao:2021vmu}.  We defer a dedicated study on any potential necessary improvements to LTCN to future work.
Partons in the CCN form as many strings as pairs of quarks present in the parton list (if that number is odd, a low $p_T$ quark is added along the beam axis). Quarks as the endpoints, and gluons as the kinks, are arranged within the different strings by minimizing the distance between two partons in $(\eta,\phi)$ space, $\Delta R=\sqrt{\Delta \phi^2+\Delta \eta^2}$ \cite{Kumar:2019bvr}. Finally, the space-time position of each of the produced hadrons, related to the break-up vertices of their associated Lund strings \cite{Ferreres-Sole:2018vgo}, is added to either $x_{\rm \tiny FO}^{\mu}$ in LTCN or the centroid of the string in CCN. For simplicity, at the present stage we ignore the negative contribution to the hypersurface energy and momentum of the sampled thermal partons.

The hadrons obtained via the Cooper-Frye procedure, from the hydrodynamized system, and those hadronized via the Lund string model, the fragmented hadrons, evolve together through the transport equations encoded in UrQMD~\cite{Bass:1998ca,Bleicher:1999xi}. Even though a study of the effect of the hadronic re-scatterings on the substructure of high-$p_T$ jets has not yet been done, there are indeed studies that demonstrate the non-negligible effects suffered by hadrons of a wide range in $p_T$~\cite{Ryu:2017qzn,Ryu:2018ckh,Dorau:2019ozd,Bierlich:2021poz}. For this reason, even the fragmented hadrons, likely coming from partons with a $p_T$ higher than average, are best described concurrently with the rest of the bulk of the system also in the hadron resonance gas phase.

\section{Results}
\label{sec:results}

In this Section we first describe the minimal adjustments needed to achieve a realistic description of experimental data on selected integrated observables in Subsection~\ref{sub:integrated}. In this way we can perform an apples to apples comparison of the most relevant features of the modified hydrodynamical evolution in Subsection~\ref{sub:modified}. We also explore more differential observables and study in detail their dependence on the minijets abundance in Subsection~\ref{sub:differential}.

\subsection{Multiplicity and $p_T$-Integrated Flow}
\label{sub:integrated}

\begin{figure}[t!]
    \centering
    \includegraphics[width=0.5\textwidth]{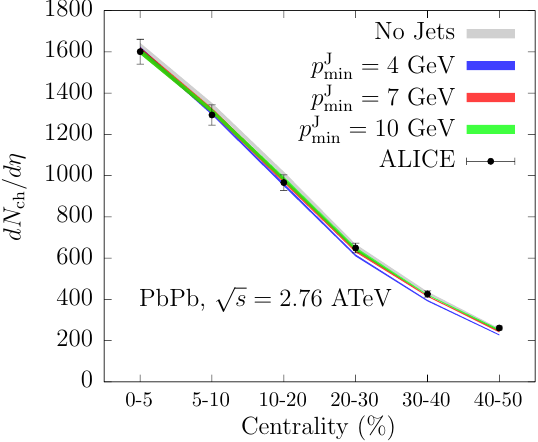}
    \caption{Charged particle multiplicity at mid-rapidity as a function of centrality for different choices of \pmin -- including ``No Jets" -- and compared to ALICE data~\cite{ALICE:2010mlf}.}
    \label{fig:multipcent}
\end{figure}

Except for a relatively small number of studies (for instance \cite{Belyaev_2020,Ryu:2018ckh,Zhao:2021vmu}), most simulations of heavy-ion collisions do not include yet the presence of the minijets. This work also features the energy loss formalism computed with holography techniques, and a pre-hydro IP-Glasma initial state.  Previous studies using the IP-Glasma+MUSIC+UrQMD framework \cite{Ryu:2017qzn, McDonald:2016vlt} therefore assumed that all the energy of a given event originates from the physics of saturation described by the CGC, effectively modeled through IP-Glasma. In this work, depending on the value of \pmin we will be including in the system new, extra, contributions to the total entropy associated to minijet production. For this reason, the amount of energy that was assumed to be contained in the low $x$, saturated modes of the CGC will necessarily have to be reduced in our new framework. We can regulate the correspondence between the energy density of the CGC and that of hydro at $\tau_0$ by using the parameter $s_{\rm factor}$.  

\begin{table}[h]
    \centering
    \begin{tabular}{c|c|c}
      \pmin  & $s_{\rm factor}$ & $\eta/ s$  \\\hline
        4 GeV & 0.45 & 0.02 \\
        7 GeV & 0.82 & 0.1 \\
        10 GeV & 0.9 & 0.125 \\ 
        No Jets & 0.915 & 0.13 \\
    \end{tabular}
    \caption{Summary of the modification of the model parameters needed to accommodate experimentally measured charged particle multiplicity and integrated flow as a function of \pmin. }
    \label{tab:newparams}
\end{table}

\begin{figure}[t!]
    \centering
    \includegraphics[width=0.5\textwidth]{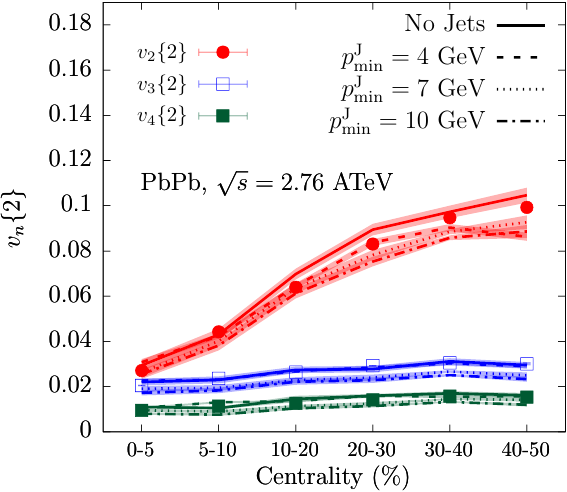}
    \caption{Integrated $v_n$ as a function of centrality for different choices of \pmin, compared to ``No Jets" and confronted against ALICE data~\cite{ALICE:2011ab}.}
    \label{fig:integratedvn}
\end{figure}

These new sources of entropy, which translate into new sources of hadron multiplicity at the end of the system evolution, require that we tune down the value of $s_{\rm factor}$ by a certain amount. We fix the new value of $s_{\rm factor}$ by comparing the mid-rapidity charged particle multiplicity from our model to experimental data from ALICE~\cite{ALICE:2010mlf}, as shown in Fig.~\ref{fig:multipcent}. For each value of \pmin, there is a single value of $s_{\rm factor}$ that allows us to describe experimental data across all centralities considered. This suggests that our assumption that a sizeable part of the final multiplicity comes from minijet production, which features an explicit $N_{\rm coll}$ scaling, receives empirical support. We show in Table~\ref{tab:newparams} the new values of $s_{\rm factor}$ needed to describe the data displayed in Figs.~\ref{fig:multipcent} and \ref{fig:integratedvn}.  As appropriate to the level of precision sought by our study, those have been obtained through a visual fit to experimental data.


\begin{figure*}[t!]
    \centering
    \includegraphics[width=1\textwidth]{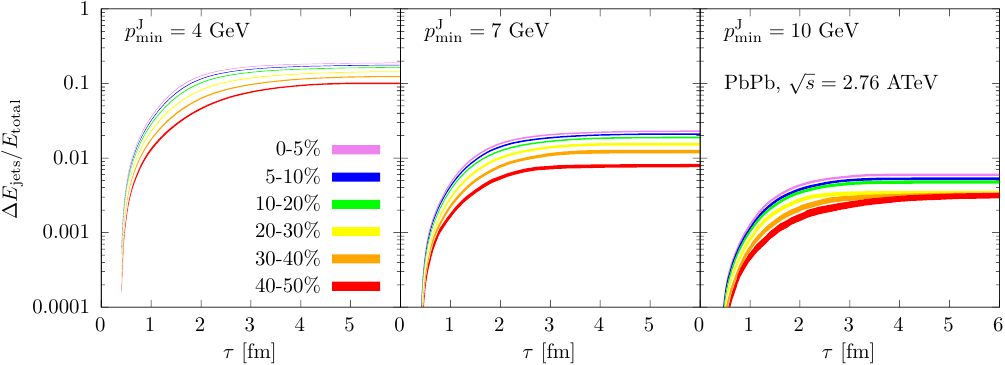}
    \caption{Fraction of the total energy contained in hydrodynamic modes injected by the mini-jets as a function of time, for different centralities and choices of \pmin.}
    \label{fig:injected}
\end{figure*}

Another important aspect of the presence of minijets in the early stages of a heavy ion collision is the reduction of collective flow \cite{Schulc:2014jma,Okai:2017ofp}. This owes to the fact that the initial orientation of the back-to-back (in the transverse plane to the beam axis) dijet pair is not correlated with the spatial anisotropies that translate into the pressure gradients driving the orientation and magnitude of collective flow. Dijet orientations are in fact random in the absence of any initial-state correlation associated with (mini-)jet production. Another reason why the magnitude of collective flow can be reduced is because the amount of energy that evolves hydrodynamically from initial time $\tau_0$ is reduced once we reduce $s_{\rm factor}$. While it is true that a sizeable amount of energy will be injected through minijet energy loss at a later time, both the delay in the injection and the fact that some of the minijet energy does not actually hydrodynamize (they escape the QGP phase) contribute into reducing collective flow.

The well-known, strong correlation between shear viscosity and the magnitude of flow, where more viscosity leads to additional entropy production and to the reduction of flow, means that the presence of minijets clearly affects the optimal value of the shear viscosity over entropy density ratio, $\eta/s$, as summarized in Table~\ref{tab:newparams}. Due to the reduction of collective flow from the minijets randomly oriented momentum deposition, one expects that $\eta/s$ needs to be reduced compared to a model without minijets in which $\eta/s$ has been adjusted to describe experimental data. We illustrate this fact in Fig.~\ref{fig:integratedvn}, where our results for the $p_T$-integrated values of $v_2$, $v_3$ and $v_4$ for different centralities, generated for different values of \pmin, are compared against experimental data from ALICE~\cite{ALICE:2011ab}. 

The introduction of mini-jets brings in non-flow effects which are suppressed in the experiments. Inspired by the experimental procedure, we calculate complex flow-vectors in two different pseudorapidity windows of 1 unit each separated by a gap of 2 units and then project one over the other. We use these final projected flow-vectors to evaluate $v_n$.

In Table~\ref{tab:newparams} we see that as we decrease \pmin, the value of $\eta/s$ necessary to reasonably describe experimental data needs to be lowered -- even below the conjectured lower bound from holography (at infinite coupling) of $\left( \eta/s \right)_{\rm AdS/CFT}=1/4 \pi$~\cite{Policastro:2001yc,Kovtun:2004de} for the case of \pmin= 4 GeV.

There certainly is a considerably larger number of relevant parameters in our model (or any other comprehensive model of heavy-ion collisions) that could be modified in order to accommodate multiplicity and $p_T$-integrated flow experimental data, potentially yielding different values of the two parameters chosen for this first exploration, $s_{\rm factor}$ and (constant, temperature independent) $\eta/s$. Nevertheless, and, as expected from the reasoning presented above, the strong variation of these two parameters suggests that they encapsulate the main distinctive physical features with respect to a model without minijets. These conjectures will need to be put to test by doing a multi-parameter fit, including as much data as possible, such as in the recent developments involving Bayesian inference techniques~\cite{Bernhard:2019bmu,JETSCAPE:2020shq,Nijs:2020ors}. 

\begin{figure*}[t!]
 \subfloat[No jets.]{\includegraphics[width=0.5\textwidth]{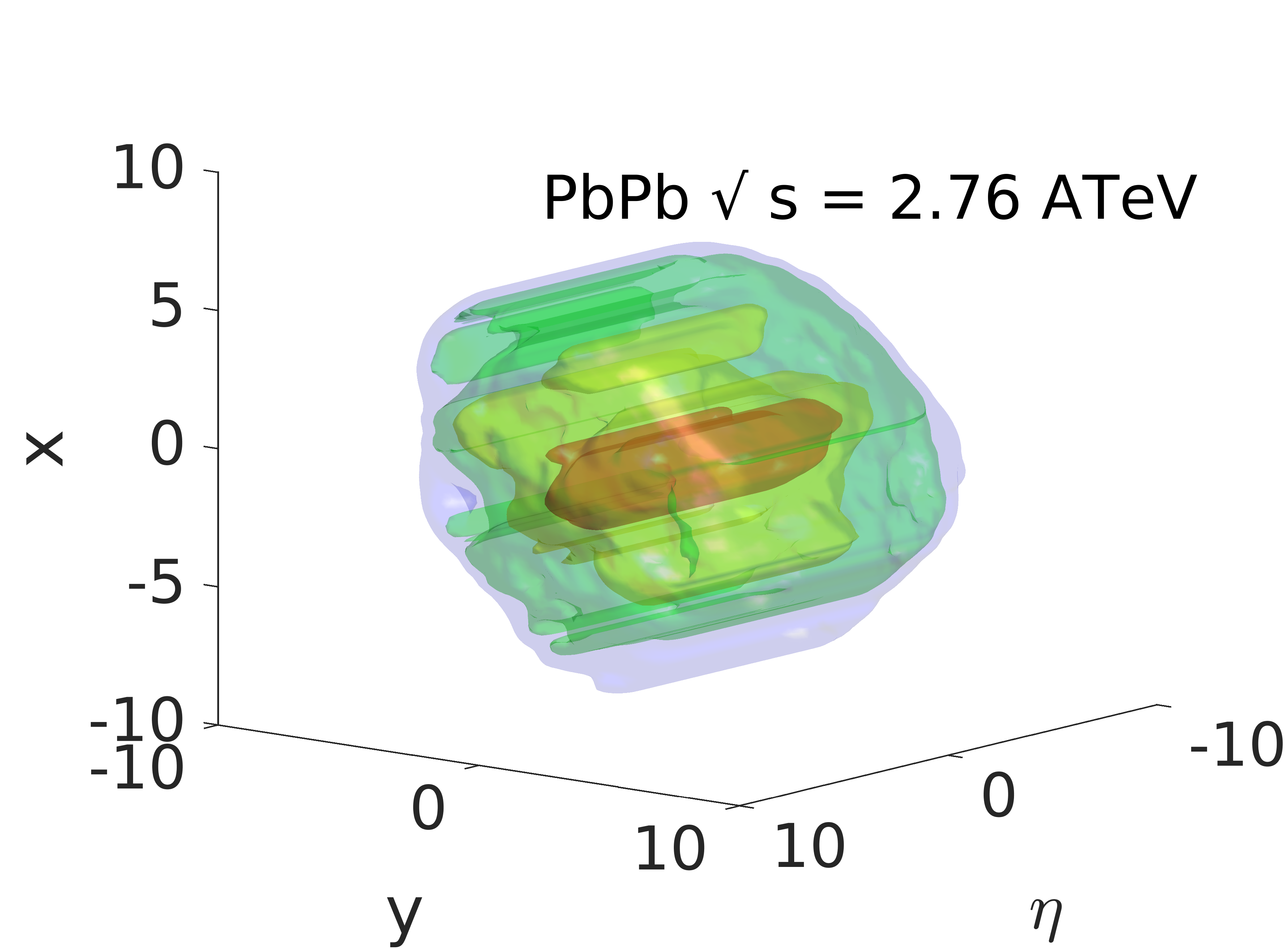}}
 \subfloat[\pmin = 10 GeV.]{\includegraphics[width=0.5\textwidth]{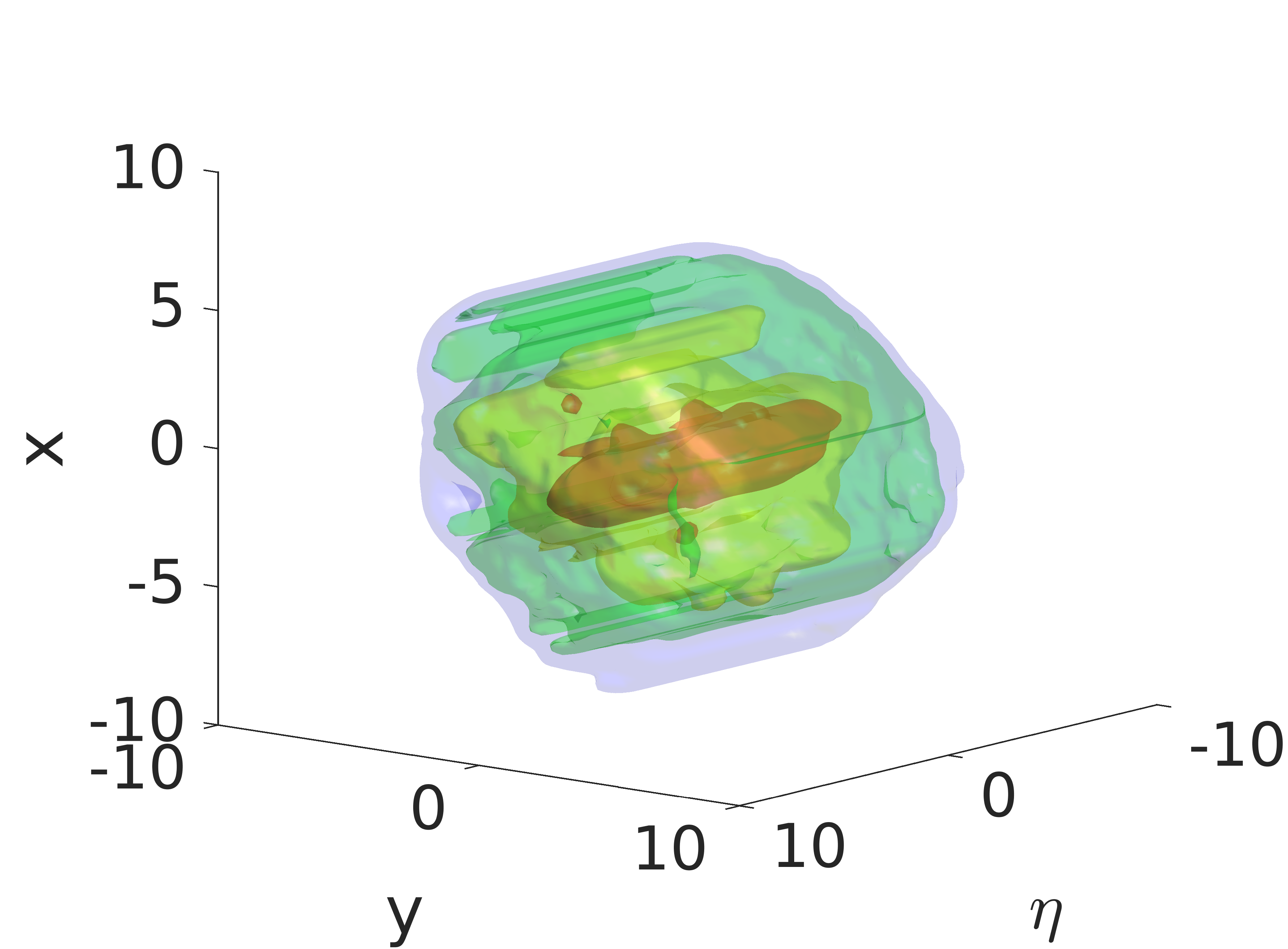}}
 \\
 \vspace{-0.3cm}
 \subfloat[\pmin = 7 GeV.]{\includegraphics[width=0.5\textwidth]{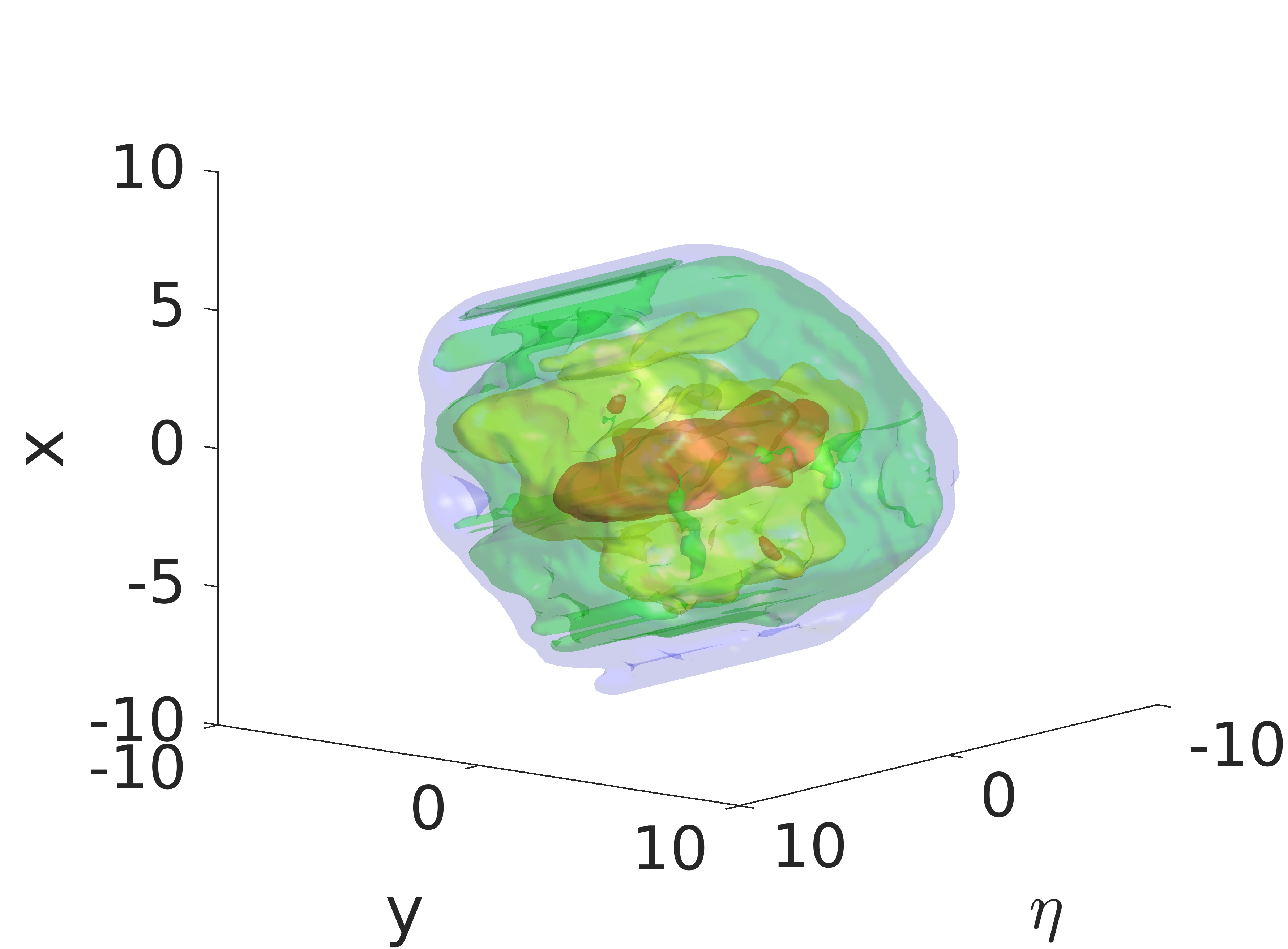}}
 \subfloat[\pmin = 4 GeV.]{\includegraphics[width=0.5\textwidth]{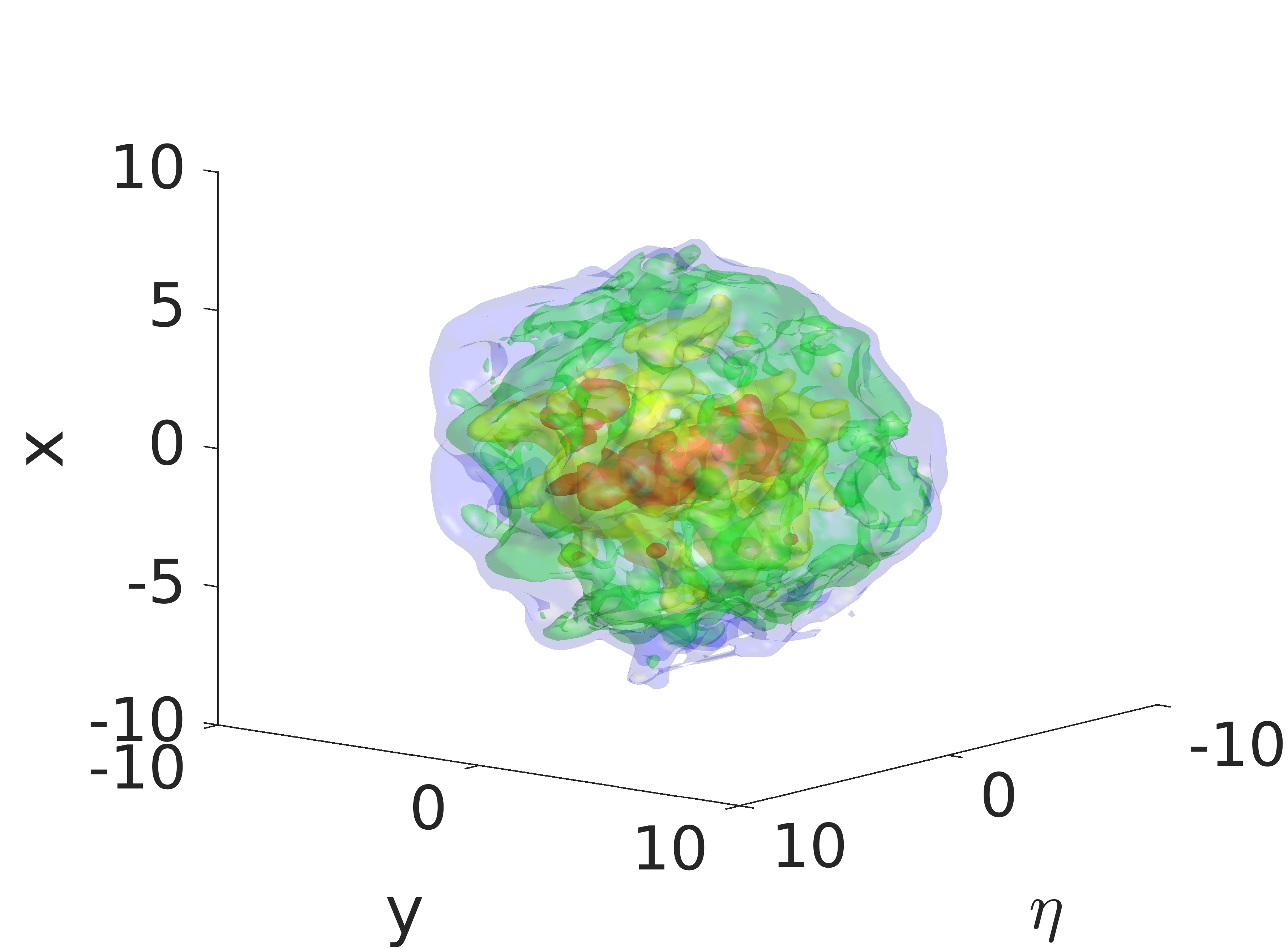}}
    \caption{Snapshots of the 3D isotherms at temperatures 220 MeV (red), 195 MeV (yellow), 170 MeV (green) and 145 MeV (blue), taken 3 fm/c after the beginning of hydro evolution, for the scenario without mini-jets in the top left panel, with \pmin = 10 GeV in the top right panel, \pmin = 7 GeV in the bottom left panel and \pmin = 4 GeV in the bottom right panel. $x$ and $y$ coordinates units are in fm.}
    \label{fig:snap}
\end{figure*}

Before moving towards a more differential study of the results obtained with this framework, it will be useful to analyze the extent to which the presence of the minijets has modified the evolution of the different stages of the system. This will lead to clues into what new phenomenological aspects to expect and where to find them. 

\subsection{A Modified Hydrodynamic Evolution}
\label{sub:modified}

\begin{figure}[t!]
  \centering
  \includegraphics[width=0.5\textwidth]{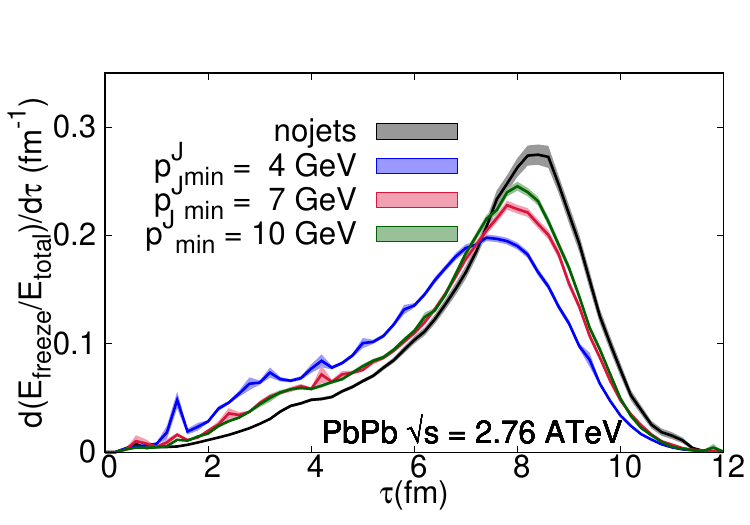}
  \caption{Fraction of energy frozen out of the 145 MeV hypersurface as a function of proper time for 30-40\% centrality bin.}
  \label{fig:freezeout}
\end{figure}

\begin{figure*}[t!]
    \centering
    \includegraphics[width=0.88\textwidth]{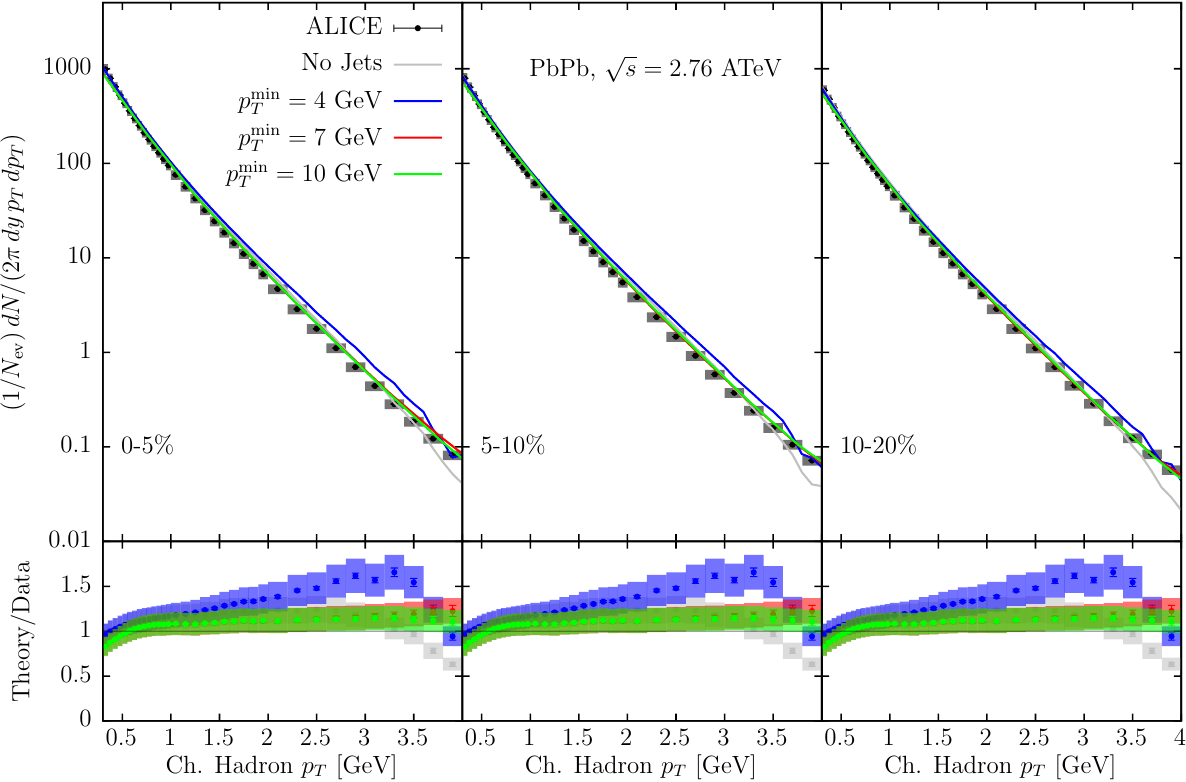}
    
    \includegraphics[width=0.88\textwidth]{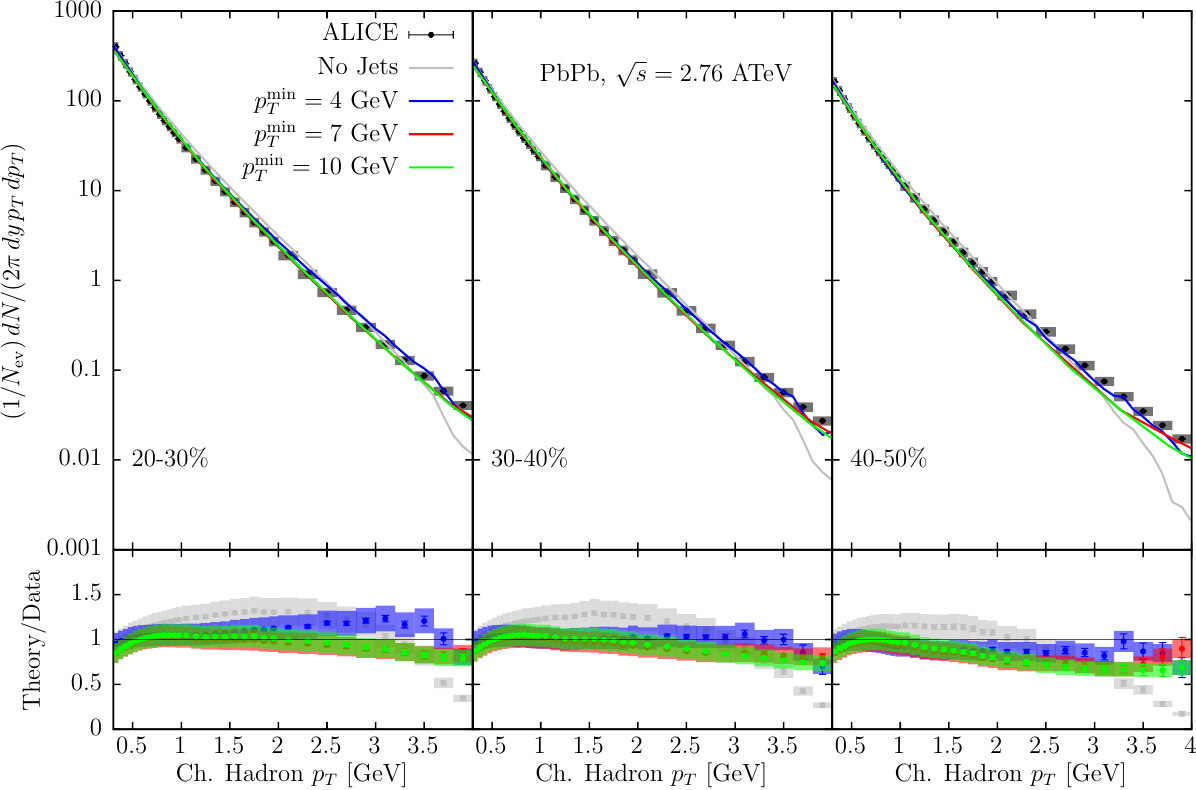}
    \caption{Charged particle $p_T$ spectrum comparing different choices of \pmin, for different centralities, confronted against ALICE data~\cite{ALICE:2012aqc}. Error bars denote statistical errors while shaded boxes correspond to the experimental systematic uncertainties.}
    \label{fig:spectranew}
\end{figure*}

The sizeable impact in multiplicity and integrated flow studied in the previous subsection calls for the modification of the amount of entropy associated to the CGC system and to the modification of the transport coefficient $\eta/s$. This is specially so for \pmin= 4 GeV, where $\eta/s$ needed to be reduced by $\approx 85 \%$. One way to understand such a large effect for this value of \pmin is by looking at the amount of injected energy from the minijets, as we do in Fig.~\ref{fig:injected}. In this Figure we show the percentage of deposited energy with respect to the total energy as a function of time $\tau$, for different centralities. Since we chose to neglect quenching effects before initial hydro time $\tau_0=0.4$ fm/c, there is no injected energy before that time. As time progresses, more and more energy is deposited from the minijets, starting to saturate around $\tau \approx 3$ fm/c, as expected from the stopping distances shown in Fig.~\ref{fig:timestop} from AdS/CFT at around $p_T \approx 4$ GeV \footnote{Note that not all partons from the minijets have to have $p_T=4$ GeV; they can be harder, and they can split into softer partons.}. For central collisions, where the average number of dijet pairs is around $\approx 500$, as shown in Fig.~\ref{fig:nhard}, we observe that close to 20 \% of the total fluid energy comes from the stopped minijets. In the most peripheral collisions, even though the number of dijet pairs is reduced by $\approx 5$, the injected energy still represents 10 \% of the total energy. Such relatively large contributions from the minijets allow us to understand the strong reduction in $\eta/s$ needed to compensate for the lessening of collective flow.

\begin{table}[t]
    \centering
    \begin{tabular}{c|c|c}
      \pmin  & $\langle N_{\rm frag.}/N_{\rm total} \rangle_{0-5\%}$ & $\langle N_{\rm frag.}/N_{\rm total} \rangle_{40-50\%}$ \\\hline
        4 GeV & 0.077(1) & 0.252(3) \\
        7 GeV & 0.0125(5) & 0.033(2) \\
        10 GeV & 0.0042(3) & 0.014(2) \\
    \end{tabular}
    \caption{The average number of fragmented hadrons (i.e. hadrons that are created due to the fragmentation of non-hydrodynamized partons) over the total number of particles prior to the evolution of the hadronic resonance gas phase, for different values of \pmin and two choices of centrality class.}
    \label{tab:fragnum}
\end{table}

The strong modifications imprinted on the hydrodynamical system can be visualized via the 3D isotherms of an event belonging to the 40-50\% centrality class shown in Fig.~\ref{fig:snap}, for different values of \pmin. 
They correspond to temperatures of 220 MeV (red), 195 MeV (yellow), 170 MeV (green) and 145 MeV (blue), taken 3 fm/c after the beginning of hydro evolution. The four panels possess the same initial profile, albeit with different values of $s_{\rm factor}$ and $\eta/s$, as specified in Table~\ref{tab:newparams}. Around this time, energy and momentum injection from the mini-jets have practically ceased, with clearly visible imprints in the different degree of spikiness and the size of the protuberances generated. These translate into local gradients that are not correlated with the system initial geometry, thereby leading to the aforementioned smaller collective flow. As expected, the profile deformation degree increases with decreasing \pmin, due to the larger abundance of mini-jet pairs. A more detailed analysis of the evolution history can be found in Appendix~\ref{app:history}. The reader will find event-averaged, energy density-weighted curves for different centralities and \pmin, as a function of time, of some relevant variables such as the temperature $T$, the transverse velocity $v_T$, the trace of the shear-stress tensor $\pi^{\mu \nu}$, the bulk pressure $\Pi$ and the momentum anisotropy $\varepsilon_p$. Even though a comprehensive study of the potentially wide phenomenological impact of the observed profile modifications is beyond the scope of the present work, from the visible differences in $v_T$ and $\pi_{\mu \nu}$ it is sensible to expect sizeable effects in low-$p_T$ photon and dilepton observables. We defer the pertinent study to a future publication.

The introduction of mini-jets also strongly influences the way fireball cools.  Mini-jets will carry matter with it in its wake breaking up high temperature isotherms.  This is essentially breaking up the QGP drop to smaller droplets which cool faster. This effect can be seen in Fig.~\ref{fig:freezeout} which shows the fraction of energy freezing out of the 145 MeV isotherm as a function of proper time. While the overall lifetime of the fireball remains about the same for different \pmin, a much larger fraction of the fireball freezes out earlier for the case with more mini-jets.

Another quantity of interest corresponds to the average number of fragmented hadrons, those that arise from the fragmentation of non-hydrodynamized partons via LTCN or CCN, with respect to the total multiplicity. We show these numbers in Table~\ref{tab:fragnum}, for different values of \pmin and two choices of centrality, the most central one and the most peripheral one used in the present work. These have been calculated before the UrQMD evolution, since only before that stage is the distinction well defined. We observe that values are larger the smaller \pmin is, as expected. For each \pmin, $\langle N_{\rm frag.}/N_{\rm total} \rangle$ increases roughly a factor $\approx 3$ by going from the central to the peripheral class. This increase is due to the fact that a smaller, colder medium will not quench mini-jets as much, decreasing the amount of hydrodynamized energy (as observed in Fig.~\ref{fig:injected}) while increasing the relative fraction of fragmented hadrons.

\begin{figure*}[t!]
    \centering
    \includegraphics[width=1\textwidth]{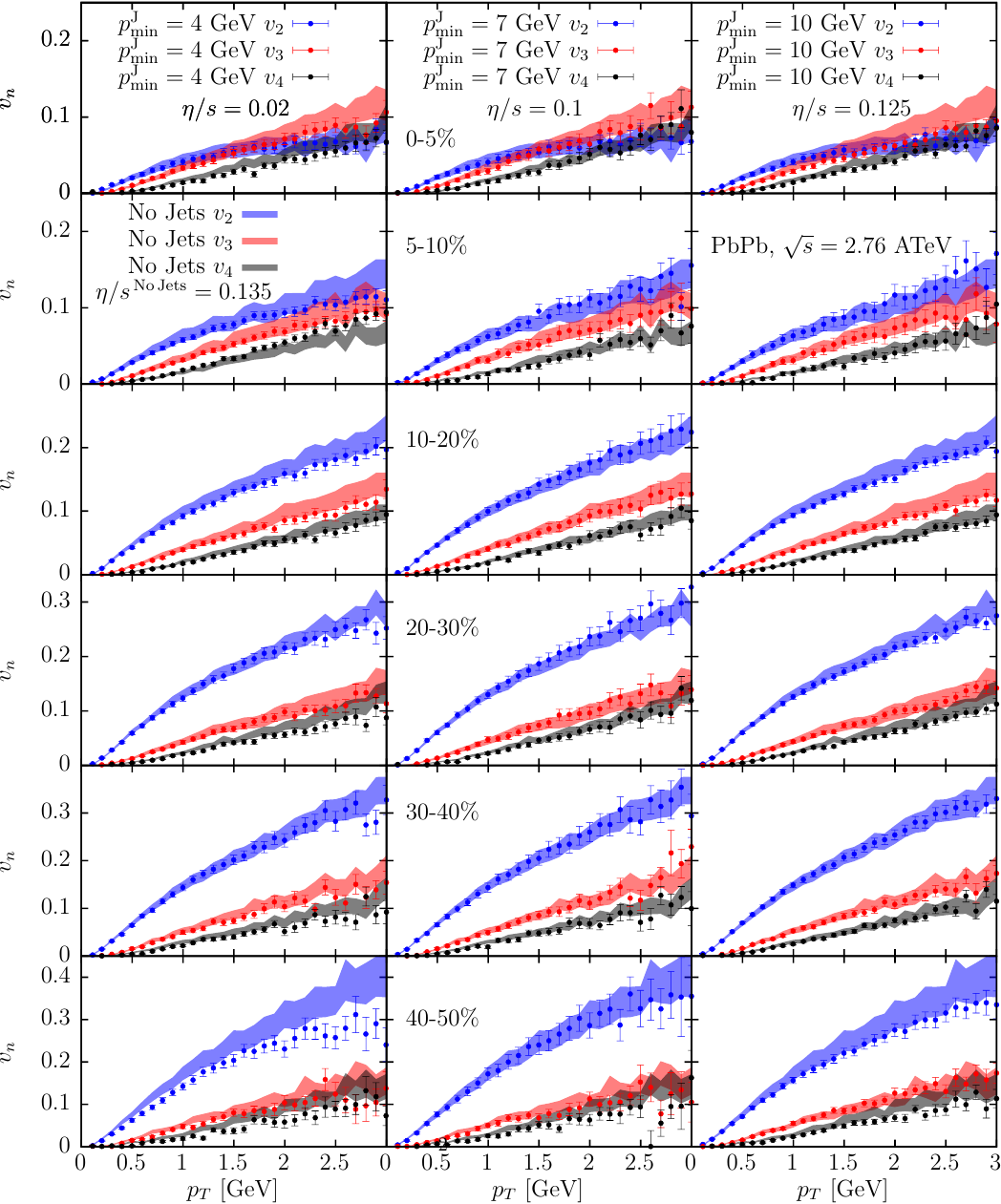}
    \caption{Flow coefficients $v_n$ as a function of charged particle $p_T$, comparing the scenario without mini-jets (solid bands) against the three different choices of \pmin (with points) displayed in the different columns. Each row corresponds to a different centrality class.}
    \label{fig:megaplot}
\end{figure*}

\subsection{Differential Observables}
\label{sub:differential}

We have seen how the inclusion of the mini-jets can be consistent with experimental data in integrated observables, such as multiplicity and integrated flow, for different centralities. We have needed to (and limited ourselves just to) adjust two of the model parameters, this is, the amount of entropy deposited in the CGC -- the $s_{\rm factor}$ -- and the specific shear viscosity, $\eta/s$. Without any further tuning of the model, we can look for other observable features that are modified employing more differential analysis.

\begin{figure*}[t!]
    \includegraphics[width=0.49\textwidth]{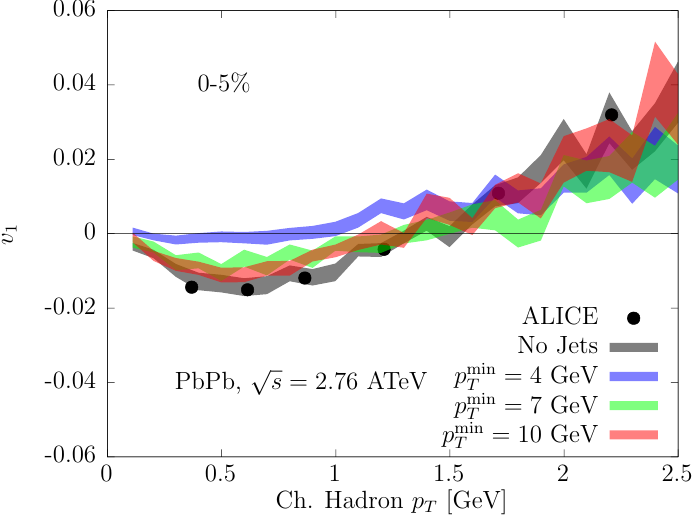}
    \includegraphics[width=0.49\textwidth]{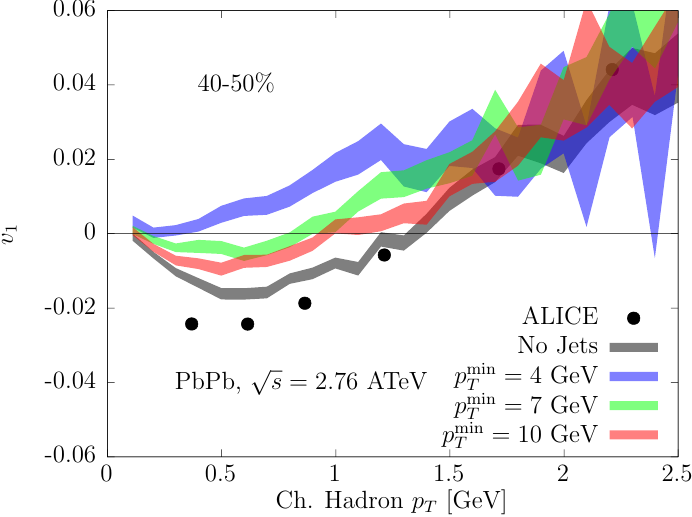}
    \caption{Directed flow $v_1$ as a function of $p_T$, comparing the ``No Jets'' scenario and the one with mini-jets with different choices of \pmin, both for central (left panel) and peripheral (right panel) centrality classes. Results are confronted against ALICE measurements, extracted using correlation data~\cite{ALICE:2011svq,Retinskaya:2012ky}.}
    \label{fig:dirflow}
\end{figure*}

The first differential observable we will focus on is charged particle production as a function of centrality, shown in Fig.~\ref{fig:spectranew}. We compare our results for the different choices of \pmin against ALICE data~\cite{ALICE:2012aqc}. 
For the three most central classes, shown in the top row of Fig.~\ref{fig:spectranew}, all scenarios, but \pmin = 4 GeV, reproduce experiments below $p_T\approx 3$ GeV. The visible overestimation of mid-$p_T$ particles with \pmin = 4 GeV, likely due to the notable contribution of fragmented hadrons, as shown in Table~\ref{tab:fragnum}, seems to disfavour this value of \pmin within our model. However, for the most peripheral class, in the bottom right panel, \pmin = 4 GeV is somewhat preferred over the other two values of \pmin. Also, as centrality decreases, we observe how all values of \pmin  \, yield a better description than the ``No Jets'' scenario below $p_T\lesssim 3$ GeV.
Above this $p_T$, ignoring the mini-jets very strongly underpredicts particle production, as expected. This disagreement is stronger with decreasing centrality. We can understand this due to the relatively larger contribution from fragmented hadrons to the total multiplicity in more peripheral centrality classes, as pointed out in Table~\ref{tab:fragnum}. It is worth noting that a single, centrality-independent readjustment of $s_{\rm factor}$ and $\eta/s$ provides a reasonable description of this observable. However, due to the relatively mild differences between the results for \pmin $>$ 4 GeV, this observable does not seem to be specially suited to discern the appropriateness of the value of the mini-jet minimum production scale in real collisions.

We now turn to the $v_n$ coefficients as a function of centrality and $p_T$, shown in Fig.~\ref{fig:megaplot}. The bands correspond to the results without mini-jets, and the dots in each column show those for different values of \pmin. For the $p_T$ range considered, and all over the analyzed centrality classes, results between different choices of \pmin and the ones without mini-jets are very similar (with the exception of the presence of a small difference for \pmin = 4 GeV in the 40-50\% centrality class). Even for the extreme case of \pmin = 4 GeV, the readjustment of $\eta/s$ necessary to get sensible values of integrated flow, as shown in Fig.~\ref{fig:integratedvn} also suffices to obtain equivalent results for $p_T$-differential flow. From these results we conclude that the destruction of collective flow caused by the mini-jets' random orientation in the transverse plane can to a large extent be compensated by a reduction of shear viscosity. At the same time, the striking similarity between the results of Fig.~\ref{fig:megaplot} means that this is not an observable capable of discriminating between different abundances of mini-jets in the initial state.

A very different picture arises with the analysis of directed flow, $v_1$, as a function of $p_T$, shown in Fig.~\ref{fig:dirflow}. $v_1$ shows a strong dependence on the presence of mini-jets, with very different results between ``No Jets'' and \pmin = 4 GeV both in the central and peripheral cases studied.
$v_1$ has been found to be approximately directly proportional to dipole asymmetry, $\epsilon_1$~\cite{Gardim:2011qn}. Dipole asymmetry is generated purely from fluctuations in the initial state. While higher harmonics probe smaller length scales, the approximate relation between $v_1 \propto \epsilon_1$ renders directed flow quite insensitive to the value of viscosity. From an hydrodynamic point of view, which the case ``No Jets'' best represents, 
the qualitative aspects of the trend of $v_1$ can be understood by noting that that high-$p_T$ particles tend to flow in the direction of the steepest gradients, while the
low-$p_T$ particles tend to flow in the opposite direction - thus the change in sign of $v_1$ as a function of $p_T$, ensuring momentum conservation. The introduction of mini-jets modifies this picture via two mechanisms. First, the injected energy and momentum produce sizeable inhomogeneities throughout the hydrodynamical system, ending up with a largely distorted dipole asymmetry by the time injection saturates around $\tau\approx 3-4$ fm/c. Second, the fragmented hadrons, those arising from partons that did not completely hydrodynamize (the corona-like contribution), have an orientation that is independent from that of the pressure gradients. In essence, both contributions together can produce a  large amount of local fluctuations in momentum space, enough to wash out the correlations associated with the initial dipole asymmetry. The effect will grow with decreasing  \pmin. Note that, despite the strong modification of the shape of $v_1$ vs. $p_T$, the momentum conservation condition, $\langle p_T\, v_1(p_T)\rangle=0$, is still preserved within statistical uncertainties for each value of \pmin.
The measured $v_1$~\cite{ALICE:2011svq,Abelev:2013cva} is very well described by hydrodynamics~\cite{Teaney:2010vd,Retinskaya:2012ky}. Strong departures from the ``No Jets'' scenario, which best describes data as shown in Fig.~\ref{fig:dirflow}, therefore imply strong tensions with experimental data, which would seem to rule out the \pmin = 4 GeV case within our model. Interestingly, this observable provides the opportunity to constrain the role of mini-jets in heavy-ion collisions, deserving special attention as well in other studies featuring non-hydrodynamic system components.

\section{Summary \& Outlook}
\label{sec:summary}
We have introduced a new framework with the capacity to evolve jets and the hydrodynamic QGP simultaneously. The energy and momentum of the mini-jets is lost to the plasma via a strongly coupled energy loss rate that depends on the local properties of the hydrodynamic system. The injection of energy and momentum through source terms in the hydrodynamic equations of motion updates the hydrodynamic profile, which is in turn affecting the mini-jets in the next time step. This type of concurrent framework is needed when dealing with a large number of such mini-jets, whose entropy represents a sizeable part of the total entropy in the system. The main goal of the current work has been the analysis of the impact of the presence of mini-jets on a limited set of well-known observables. The abundance of mini-jets, which have a finite probability to be produced at each binary collision, is greatly determined by the allowed minimum $p_T$, \pmin, that the corresponding inelastic process can have. We have used three different values, \pmin= 4, 7, 10 GeV, all of them larger than the saturation scale $Q_s\approx 2$ GeV as we assume that mini-jet production is decoupled from the low-$x$ physics responsible for the evolution of the saturated Glasma state.

In order to keep total multiplicity within the experimentally measured range, the entropy associated with the initial state needs to be reduced to compensate for the extra entropy contributed by the mini-jets. This is done by rescaling the parameter $s_{\rm factor}$ when the Glasma system is matched to the hydrodynamic stress-energy tensor at hydrodynamic initialization time, $\tau_0=0.4$ fm/c. Before this time, mini-jets evolve according to a space-time picture based on formation time arguments, without interacting. After $\tau_0$, partons can inject energy and momentum to the plasma, a process that typically saturates around $\tau \approx 3 - 4$ fm/c. The fact that this time is considerably larger than other commonly used hydrodynamization times, closer to 1 fm/c~\cite{Kurkela:2018wud,Gale:2022hld}, means that there are parts of the system that will not hydrodynamize at the same pace than the rest of the bulk, rendering a single hydrodynamization time insufficient to describe all the relevant non-equilibrium dynamics.

Even more importantly, the fact that the orientation of the mini-jets is uncorrelated with the direction of the initial pressure gradients leads to a destruction of collective flow. In order to restore the experimentally acceptable values for the integrated flow coefficients, $v_n$, one needs to reduce viscosity, most notably $\eta/s$. Transport coefficients have the power to provide information about the microscopic nature of the QGP~\cite{Kovtun:2004de,Ghiglieri:2018dib}, and their extraction from model comparison to data is still under active investigation~\cite{Bernhard:2019bmu,JETSCAPE:2020shq,Nijs:2020ors}. From the results obtained in this work, it is clear that the physics of mini-jets ought to be included in the models used in such parameter extraction exercises.

A single rescaling of the $s_{\rm factor}$ and of $\eta/s$ values per choice of \pmin \, suffices to describe with reasonable accuracy the measured values for multiplicity and integrated $v_n$ across a wide range of centrality classes. For the limited $p_T$ range studied in this work, with $p_T<3$ GeV, differential $v_n$, with $2\leq n \leq 4$, does not discriminate between different values of \pmin. Despite the important system modifications introduced by the presence of the mini-jets, an appropriate reduction of $\eta/s$ can equalize the $p_T$-differential flow strength among the different scenarios. In stark contrast lie the results for directed flow, $v_1$. This observable \emph{is} very sensitive to the presence of the mini-jets, and has the potential to constrain the relevant scales and initial stages related to mini-jet production in heavy-ion collisions.

We find it appropriate to emphasize the exploratory nature of our present study; we have simply adjusted two among the many parameters present in this comprehensive model of heavy-ion collisions.  A meaningful,  strong conclusion about the eventual necessity to modify the underlying physical phenomena used in this work would require a complete scan of such many parameters,  for instance allowing for a more general parametrization of the functional dependence of $\eta/s$ with temperature,  taken to be simply a constant for the moment. Such a parameter-scan could be incorporated in holistic Bayesian studies such as those in Refs.  \cite{Bernhard:2019bmu,JETSCAPE:2020shq,Nijs:2020ors}.

The consequences of the introduction of this new element in the Standard Model of heavy-ion collisions, the mini-jets, are numerous and far-reaching. We provide a non-exhaustive list of the studies needed to better understand them:
\begin{itemize}
    \item Study other center-of-mass energies, in particular the lower ones at RHIC. The important differences in the jet spectrum can put to test the consistency of the framework with respect to the higher center-of-mass energies here studied.
    \item We have limited ourselves to a given parton energy loss model. One should expect that different energy loss models could yield different results. This suggests that the physics of parton, or jet, energy loss can be constrained by the analysis of bulk observables, and not only by the high-$p_T$ jet observables.
    \item The hydrodynamic profile can be substantially modified depending on the choice of \pmin. It would be interesting to analyze how this affects high-$p_T$ jet quenching observables, given in particular the sizeable new fluctuations introduced by the mini-jets orientation, or the presence of a delay ($\approx 3-4$ fm/c) in the hydrodynamization of a sizeable part of the total energy and momentum of the system.
    \item Due to their widely varying rapidities, mini-jets introduce a new source of fluctuations that should impact event-plane decorrelation with rapidity, also called $r_n$~\cite{Bozek:2010vz,Schenke:2016ksl,Pang:2015zrq}. 
    \item The modified hydrodynamic profile can also have an impact on photon observables. Photons represent clean probes, emitted throughout the system evolution and barely re-scattering~\cite{Paquet:2015lta}. We expect them to be sensitive to the modification of the evolution history induced by the mini-jets.
    \item The study of results at higher $p_T>3$ GeV, which will allow to check the way in which the low-$p_T$ observables match the high-$p_T$ ones. The intermediate $p_T$ region can be sensitive to quark coalescence dynamics~\cite{Zhao:2021vmu}, an hadronization mechanism that is described only approximately in our current framework via LTCN.
\end{itemize}

From the theoretical point of view, having a complete description of the initial stages capable of accounting for the mid-$p_T$ mini-jet objects along with the saturated Glasma is clearly the most pressing goal. Recent findings on the largeness of the jet quenching parameter $\widehat{q}$ in the Glasma~\cite{Ipp:2020nfu,Ipp:2020mjc,Carrington:2021dvw,Carrington:2022bnv} provide a strong motivation to include these physics in future model improvements. Such efforts will contribute to reduce the divergence in the modelling assumptions present among current comparable concurrent jet+hydro framework studies~\cite{Lokhtin:2008xi,Werner:2013tya,Okai:2017ofp,Chen:2017zte,Karpenko:2019xsc,Zhao:2021vmu,Kanakubo:2021qcw}, leading to a more robust extraction of the QGP transport coefficients and a better understanding of the physics of hydrodynamization of deconfined QCD matter.  Importantly,  this work demonstrates the intricate interplay between mini-jet energy loss and the hydrodynamical evolution of the QGP.

\acknowledgements
We acknowledge useful discussions with Krishna Rajagopal, Wilke van der Schee, Soeren Schlichting, Chun Shen, Sangwook Ryu, Yasuki Tachibana and Konrad Tywoniuk. This work was funded in part by the Natural Sciences and Engineering Research Council of Canada. Computations were made on the B\'eluga supercomputer at McGill University, managed by Calcul Qu\'ebec and Compute Canada.
D.P. has received funding from the European Union’s Horizon 2020 research and innovation program under the Marie Skłodowska-Curie grant agreement No. 754496. M.S. is also supported by U.S. DOE Grant No. DE-FG02-87ER40328. 

\bibliography{biblio}

\appendix

\section{Analysis of hydrodynamical evolution history}
\label{app:history}
In this Section we present the modification of the evolution of some hydrodynamic quantities, averaged over a few events ($\approx$ 30), due to the presence of the mini-jets, for different values of \pmin, across several centralities.

Each quantity $X$ is averaged for a given $\tau$ across all the volume with $T>T_c$ in the event $i$, weighted by the energy density $\varepsilon$ as
\begin{equation}
    \langle X \rangle_i \equiv \frac{\int dV X_i(x) \varepsilon_i (x)}{\int dV \varepsilon_i(x)} \, .
\end{equation}
Then, the final event-averaged quantity, denoted for simplicity just as $\langle X \rangle$, without subscript $i$, is also weighted according to the total energy of event $i$, as
\begin{equation}
    \langle X \rangle \equiv \frac{\sum_i \langle X \rangle _i E_{i}}{\sum_i E_{i}} \, ,
\end{equation}
where $E_i\equiv \int dV \varepsilon_i(x)$.

\begin{figure*}[t!]
    \centering
    \includegraphics[width=0.88\textwidth]{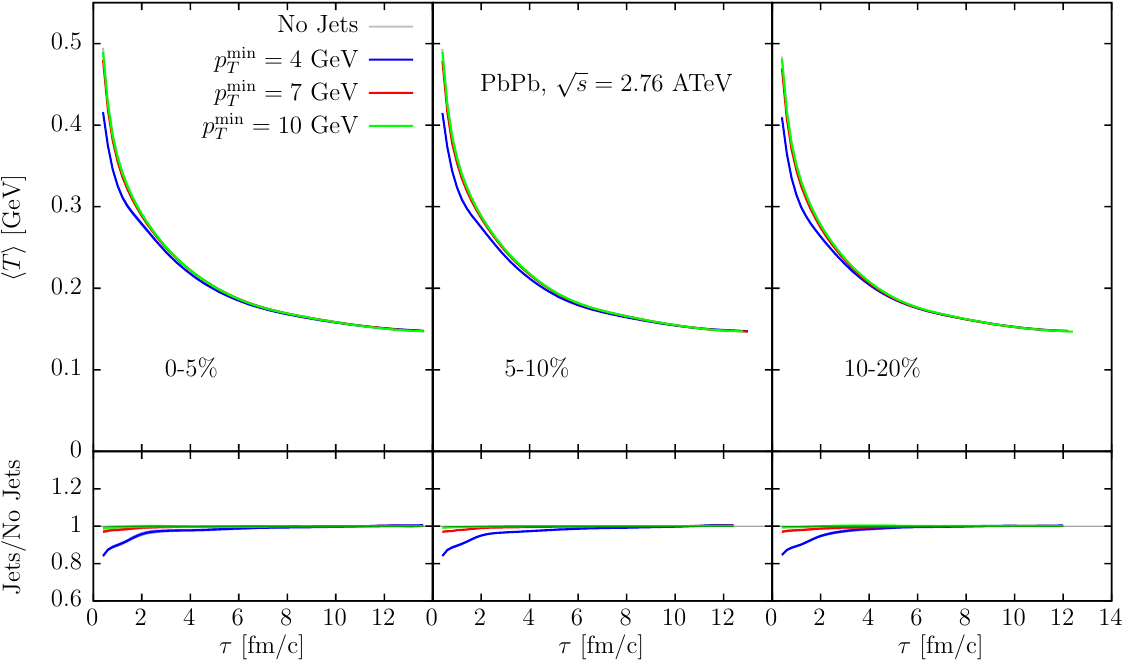}
    
    \includegraphics[width=0.88\textwidth]{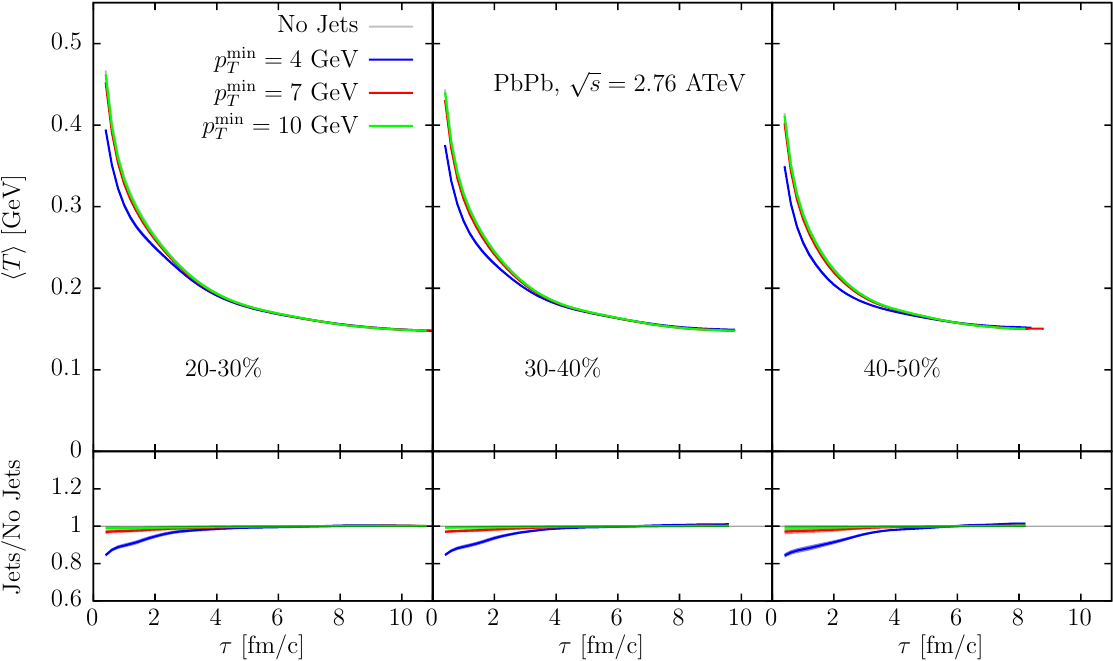}
    \caption{Time evolution of the energy density-weighted average temperature $\langle T \rangle$, for different values of \pmin and several centrality classes.}
    \label{fig:temp_history}
\end{figure*}

\begin{figure*}[t!]
    \centering
    \includegraphics[width=0.88\textwidth]{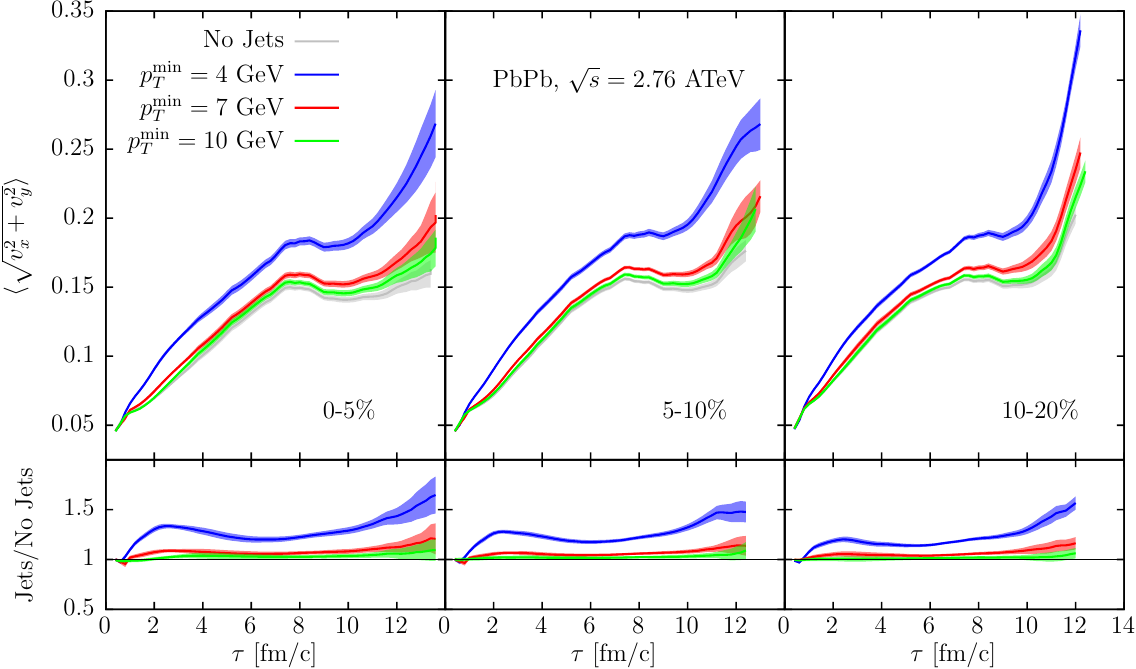}
    
    \includegraphics[width=0.88\textwidth]{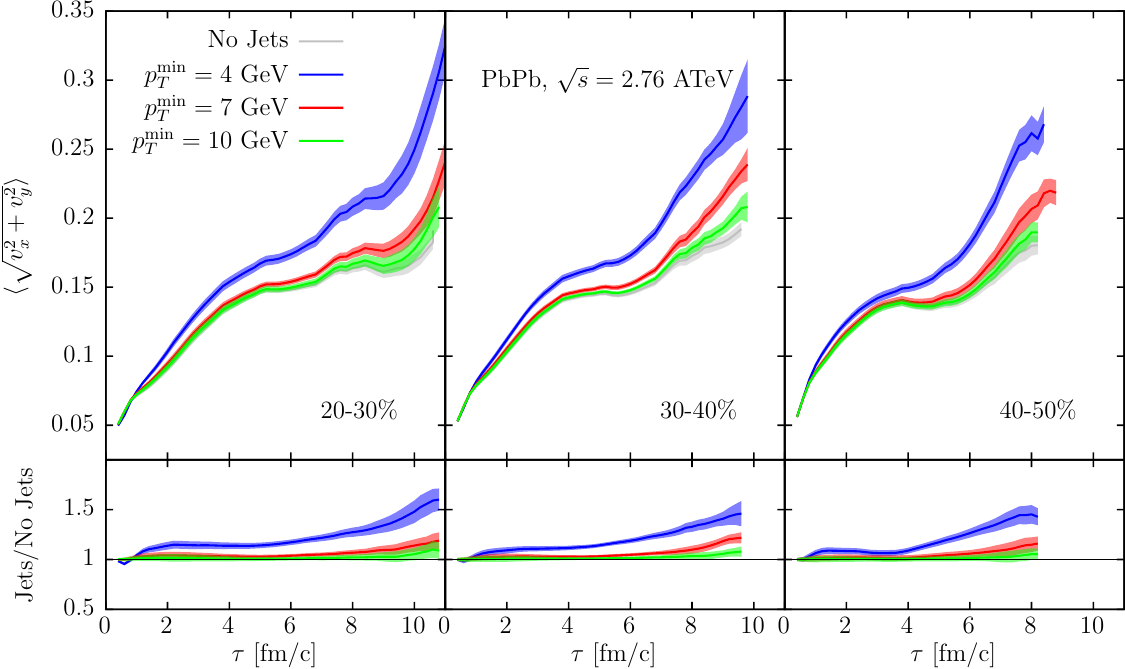}
    \caption{Time evolution of the energy density-weighted average transverse velocity $\langle \sqrt{v_x^2+v_y^2}\rangle$, for different values of \pmin and several centrality classes.}
    \label{fig:vt_history}
\end{figure*}

\begin{figure*}[t!]
    \centering
    \includegraphics[width=0.88\textwidth]{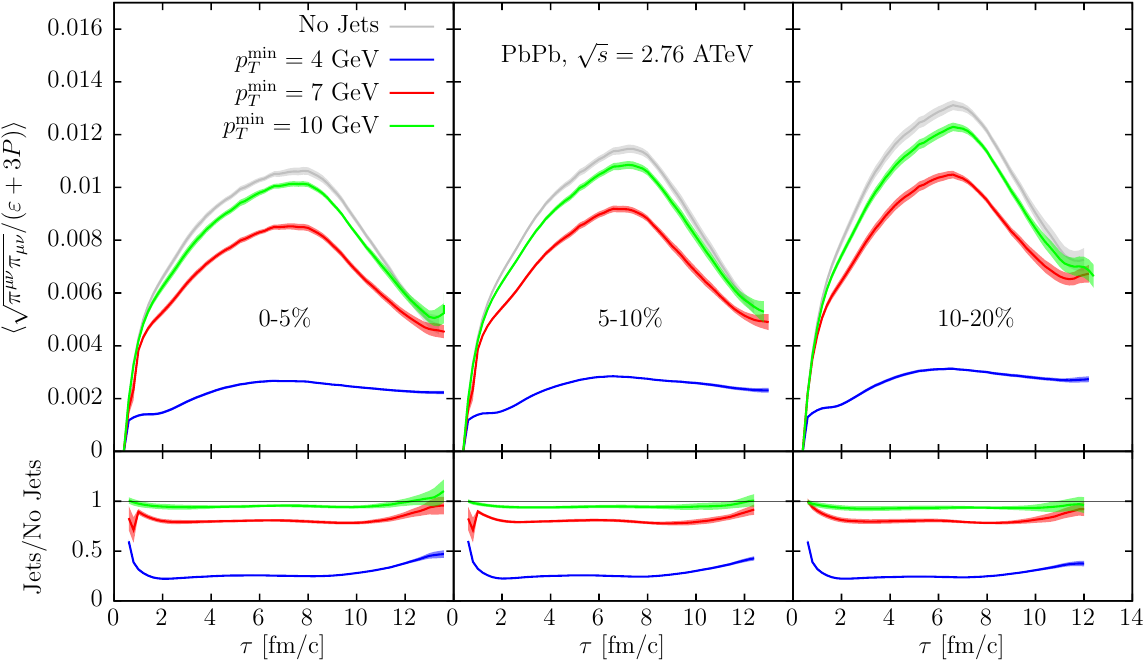}
    
    \includegraphics[width=0.88\textwidth]{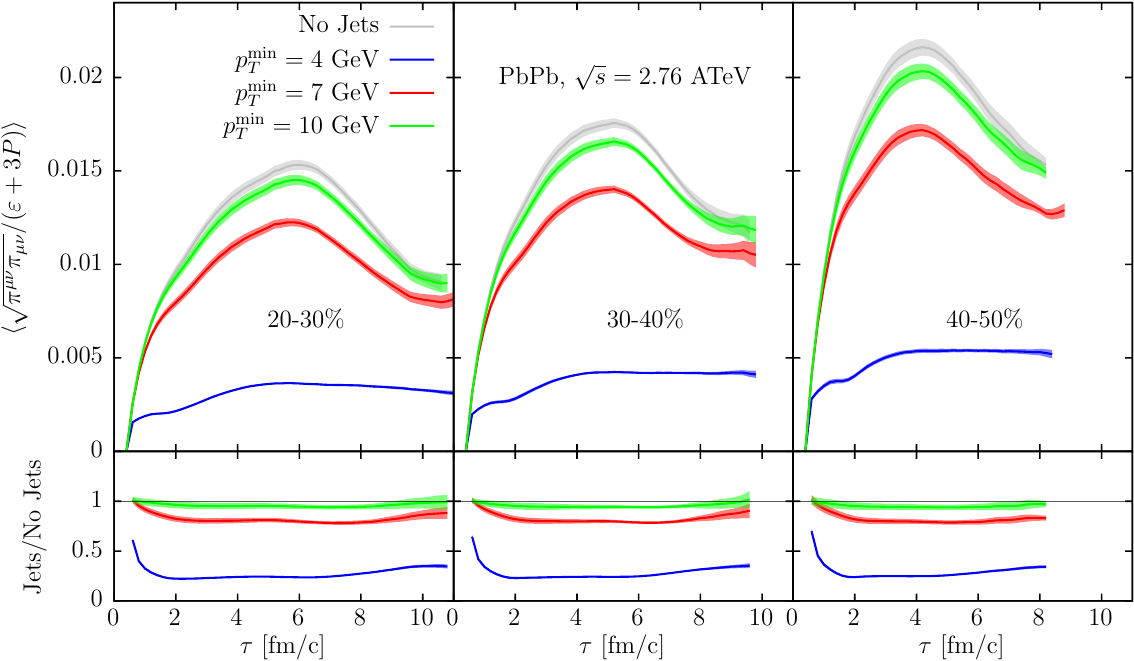}
    \caption{Time evolution of the energy density-weighted average trace of the shear stress tensor, normalized by the trace of the ideal stress tensor, $\langle \sqrt{\pi^{\mu \nu}\pi_{\mu \nu}}/(\varepsilon+3P) \rangle$, for different values of \pmin and several centrality classes.}
    \label{fig:shear_history}
\end{figure*}

\begin{figure*}[t!]
    \centering
    \includegraphics[width=0.88\textwidth]{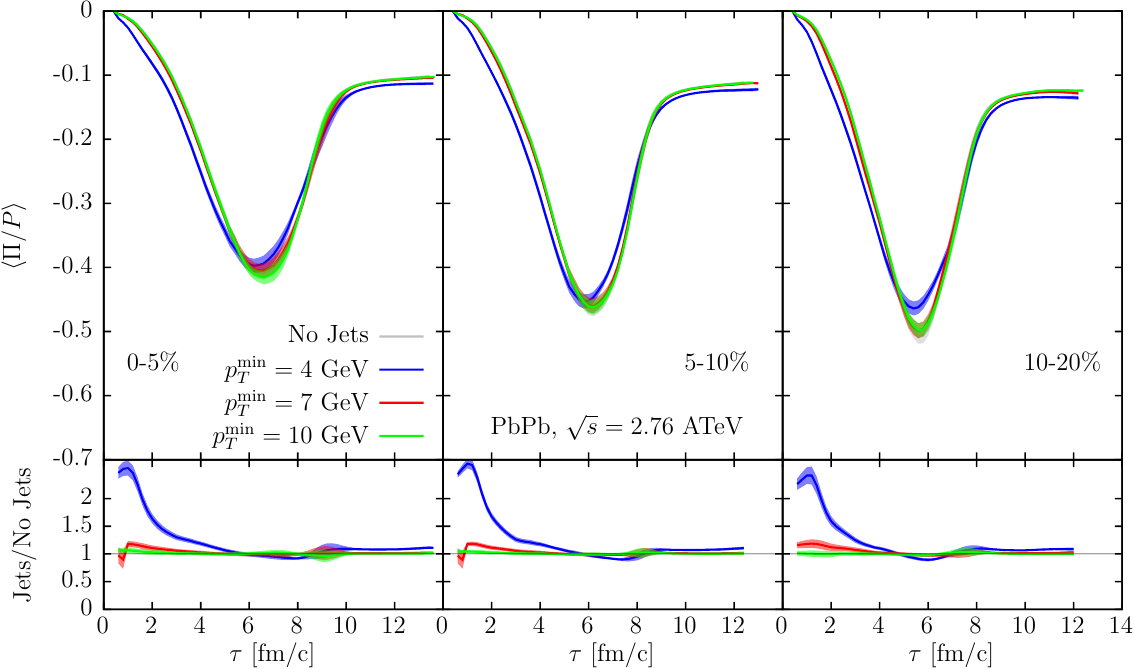}
    
    \includegraphics[width=0.88\textwidth]{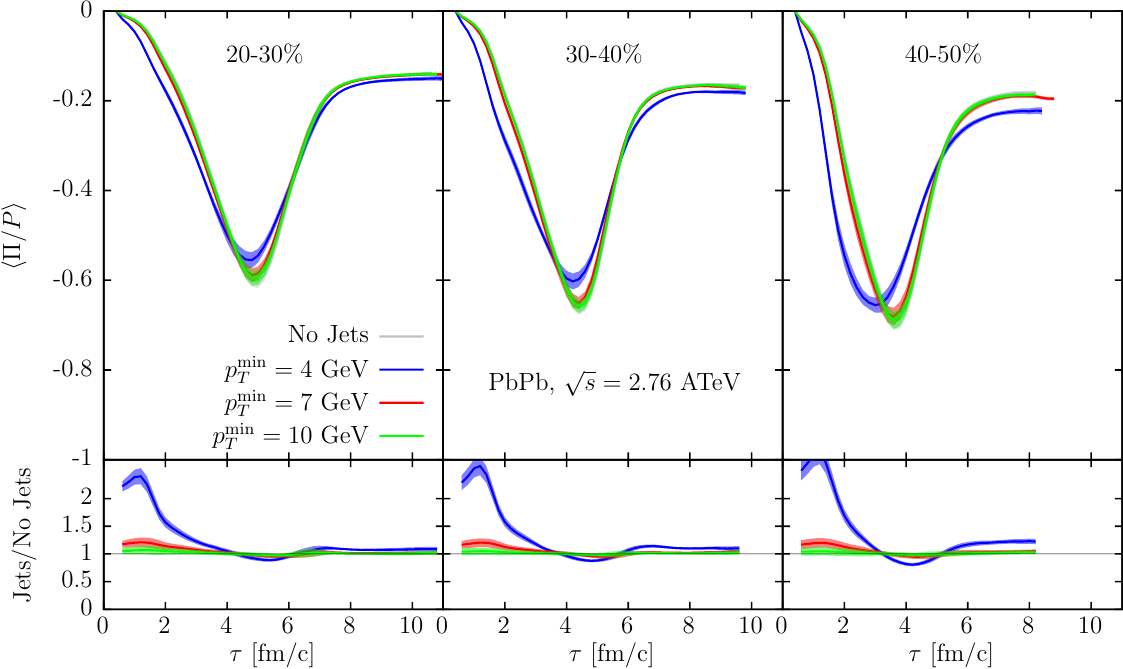}
    \caption{Time evolution of the energy density-weighted average bulk viscous pressure, normalized by the thermodynamic pressure, $\langle \Pi/P \rangle$, for different values of \pmin and several centrality classes.}
    \label{fig:bulk_history}
\end{figure*}

\begin{figure*}[t!]
    \centering
    \includegraphics[width=0.88\textwidth]{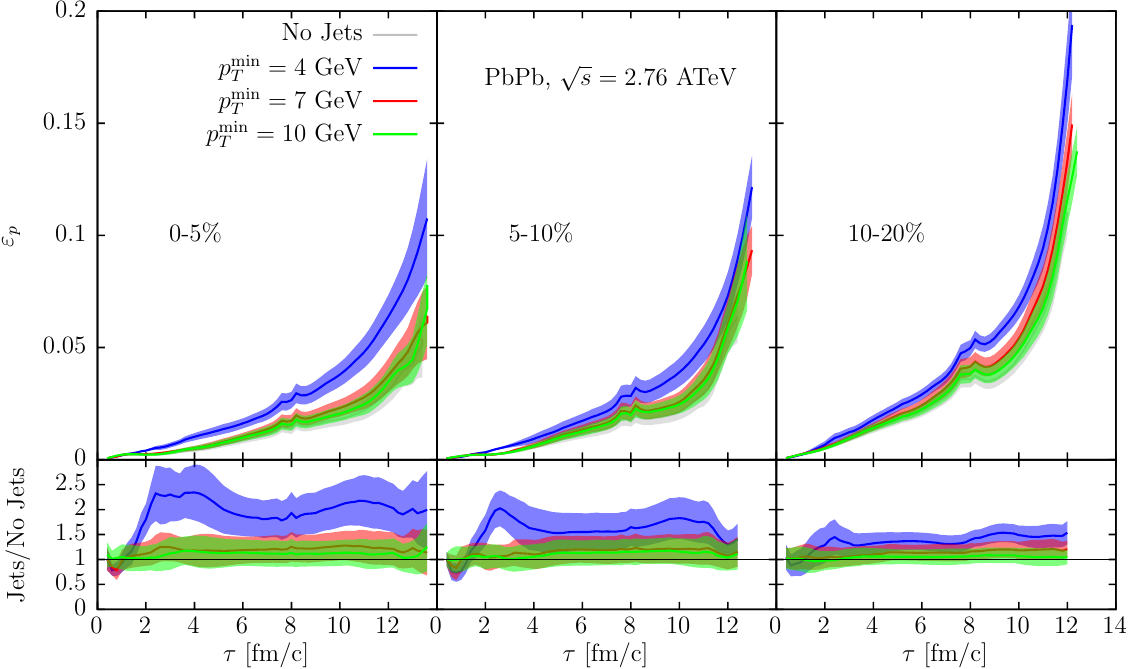}
    
    \includegraphics[width=0.88\textwidth]{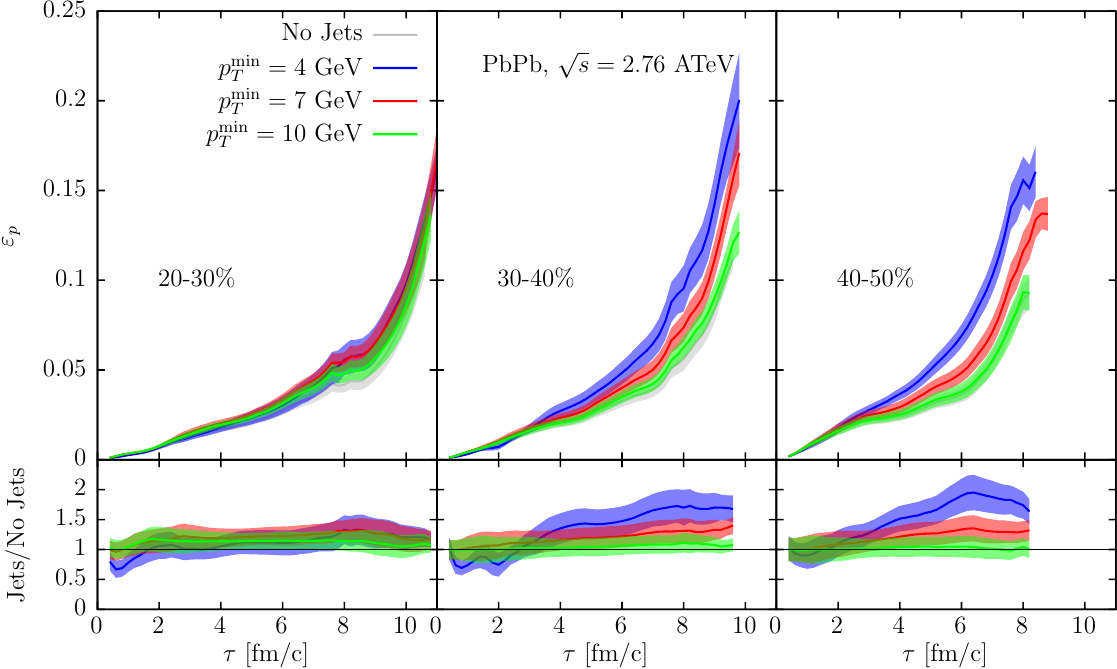}
    \caption{Time evolution of the energy density-weighted average momentum anisotropy $\varepsilon_p$, for different values of \pmin and several centrality classes.}
    \label{fig:momanip_history}
\end{figure*}

In Fig.~\ref{fig:temp_history} we see how only for the case of \pmin = 4 GeV is the temperature initially visibly lower than the one without mini-jets, since it is the only scenario for which a large part of the total energy of the system is injected by the mini-jets. The delay observed is consistent with the estimates of Fig.~\ref{fig:timestop} and Fig.~\ref{fig:injected}. The transverse velocity shown in Fig.~\ref{fig:vt_history}, increasing with decreasing \pmin, is strongly correlated with the associated decrease of the shear stress tensor with decreasing \pmin, as shown in Fig.~\ref{fig:shear_history}. Strongest deviations in the ratio of the bulk viscous pressure over the thermodynamic pressure, shown in Fig.~\ref{fig:bulk_history}, are again for the \pmin = 4 GeV case at early times, basically due to the initially reduced thermodynamic pressure, related to the delay in the rising of temperature, as shown in Fig.~\ref{fig:temp_history} (recall that the bulk viscosity parameter has been chosen to be unmodified in the present work). Finally, we show momentum anisotropy $\varepsilon_p$, defined for a given event $i$ as
\begin{equation}
    \varepsilon_{p,i} \equiv \frac{\sqrt{\langle (T^{xx}-T^{yy})\rangle_i^2+4 \langle T^{xy}\rangle_i^2}}{\langle T^{xx}+T^{yy}\rangle_i} \, ,
\end{equation}
where $T^{\mu\nu}$ is the full stress-energy tensor of the system, in Fig.~\ref{fig:momanip_history}.  Reducing $\eta/s$ (as one decreases \pmin) will tend to increase the momentum anisotropy $\varepsilon_p$ of the bulk of the system,  as can be seen most clearly in the most central panels of Fig.~\ref{fig:momanip_history}.  We note that even though $\varepsilon_p$ is much larger for smaller \pmin, the integrated $v_n$ in Fig.~\ref{fig:integratedvn} is about the same.  This is because a larger fraction of the system energy freezes out much sooner when more mini-jets are present, as seen in Fig.~\ref{fig:freezeout}.

These modifications, largest in the case of \pmin = 4 GeV, although clearly non-neglibile for \pmin = 7 GeV, motivate a more detailed study, possibly within the context of the phenomenological impact on electromagnetic probes.

\end{document}